\pdfoutput=1
\newcommand*{\ATLASLATEXPATH}{}
\documentclass[PAPER, atlasdraft=true, texlive=2016, UKenglish, cernpreprint]{\ATLASLATEXPATH atlasdoc}

\usepackage{\ATLASLATEXPATH atlaspackage}
\usepackage{\ATLASLATEXPATH atlasheavyion}
\usepackage{\ATLASLATEXPATH atlasmisc}
\usepackage{\ATLASLATEXPATH atlasbiblatex}
\usepackage{\ATLASLATEXPATH atlascontribute}
 
\usepackage{\ATLASLATEXPATH atlasphysics}
 
\graphicspath{{logos/}{figures/}}
 
\usepackage{ANA-HION-2018-25-PAPER-defs}

\AtlasTitle{Measurement of $W^\pm$ boson production in Pb+Pb collisions at $\sqrt{s_\mathrm{NN}} = 5.02$~{Te\kern -0.1em V} with the ATLAS detector}
 
\AtlasAbstract{
A measurement of $W^\pm$ boson production in Pb+Pb collisions at $\sqrt{s_\mathrm{NN}} = 5.02$~{Te\kern -0.1em V} is reported using data recorded by the ATLAS experiment at the LHC in 2015, corresponding to a total integrated luminosity of $0.49\;\mathrm{nb^{-1}}$. The $W^\pm$ bosons are reconstructed in the electron or muon leptonic decay channels. Production yields of leptonically decaying $W^\pm$ bosons, normalised by the total number of minimum-bias events and the nuclear thickness function, are measured within a fiducial region defined by the detector acceptance and the main kinematic requirements. These normalised yields are measured separately for $W^+$ and $W^-$ bosons, and are presented as a function of the absolute value of pseudorapidity of the charged lepton and of the collision centrality. The lepton charge asymmetry is also measured as a function of the absolute value of lepton pseudorapidity. In addition, nuclear modification factors are calculated using the $W^\pm$ boson production cross-sections measured in $pp$ collisions. The results are compared with predictions based on next-to-leading-order calculations with CT14 parton distribution functions as well as with predictions obtained with the EPPS16 and nCTEQ15 nuclear parton distribution functions. No dependence of normalised production yields on centrality and a good agreement with predictions are observed for mid-central and central collisions. For peripheral collisions, the data agree with predictions within 1.7 (0.9) standard deviations for $W^-$ ($W^+$) bosons.
}
 
\author{The ATLAS Collaboration}
\PreprintIdNumber{CERN-EP-2019-123}
 
\AtlasRefCode{HION-2018-25}

\AtlasJournalRef{Eur. Phys. J. C 79 (2019) 935}
\AtlasDOI{10.1140/epjc/s10052-019-7439-3}
 
\hypersetup{pdftitle={ATLAS document},pdfauthor={The ATLAS Collaboration}}
 
\begin{document}
 
\maketitle
 
% The next lines are included from the .//Sec_1.tex input file
\section{Introduction}
\label{sec:intro}
Collisions of lead ions in the Large Hadron Collider (LHC) allow the formation of a hot and dense medium with temperatures significantly exceeding the critical temperature for a phase transition from ordinary to strongly interacting matter~\cite{Shuryak:2014zxa}. Experiments at the Relativistic Heavy Ion Collider~(RHIC) at lower energies than the LHC established that strongly interacting matter takes the form of a quark--gluon plasma~(QGP)~\cite{Arsene:2004fa,Back:2004je,Adams:2005dq,Adcox:2004mh}. Particles carrying colour charge produced in hard interactions of quarks and gluons in the initial stages of a nuclear collision lose energy while traversing the QGP, leading to the phenomenon of jet quenching~\cite{Mehtar-Tani:2013pia,Qin:2015srf}. This phenomenon was established by the observation of the suppression of charged-hadron yields in heavy-ion collisions, which was reported by experiments both at RHIC and the LHC, see e.g. Refs.~\cite{Adams:2003kv,Adler:2003au,HION-2011-03,CMS-HIN-10-005,Aamodt:2010jd}. Other studies related to jet quenching include LHC measurements of the suppression of inclusive jet yields~\cite{HION-2017-10, CMS-HIN-13-005}, dijet transverse momentum imbalance~\cite{HION-2010-02, CMS-HIN-11-013} and modifications to the jet fragmentation~\cite{HION-2017-04, CMS-HIN-16-020}.
 
At all stages of QGP evolution, colourless elementary particles created in hard scatterings are expected to interact only weakly with the medium, which makes them excellent probes of the very initial stage of the collision. Moreover, in heavy-ion collisions energetic particles are produced in the interaction between nucleons in the nuclei. The latter are complex objects, so the geometry of the collision plays a central role in the interpretation of the experimental results.
The RHIC experiments measured the properties of highly energetic (virtual) photons~\cite{STAR:2016use,Afanasiev:2012dg} in Au+Au collisions and found that their production rates scale with the nuclear thickness. At the LHC, the ATLAS and CMS experiments measured the production of isolated prompt photons~\cite{HION-2012-02, CMS-HIN-11-002}, $Z$~\cite{HION-2012-07, CMS-HIN-10-003} and $W^\pm$ bosons~\cite{HION-2013-05,CMS-HIN-11-008} in lead--lead~(Pb+Pb) collisions at $\sqrtNN = 2.76$~\TeV. In addition, the forward production of $Z$ bosons was measured by the ALICE experiment in \PbPb\ collisions at $\sqrtNN = 5.02$~\TeV~\cite{Acharya:2017wpf}. The production rates of electroweak~(EW) vector bosons were found to be unaffected by the presence of the QGP, and in agreement with the expectations from the collision geometry.
 
Production of EW bosons is an important benchmark process at hadron colliders. Measurements in proton--proton~(\pp) collisions at $\sqs = 5.02$, 7, 8 and 13~\TeV\ at the LHC~\cite{HION-2018-02-witherratum, STDM-2012-20, ATLAS_W8TeV, STDM-2015-03, CMS-EWK-10-005, CMS-SMP-12-011-correct} and at previous colliders at lower energies~\cite{Abulencia:2005ix, Abazov:2013dsa} are well described by calculations based on higher-order perturbative quantum chromodynamics~(QCD) and the theory of EW interactions.
At  leading order, $W^\pm$ bosons are preferentially produced in $u\bar{d} \rightarrow W^+$ and $d\bar{u}\rightarrow W^-$ processes~\cite{Martin:1999ww}.
In \PbPb\ collisions, due to the different proportions of $u$ and $d$ quarks in the proton compared to the lead nucleus, the individual $W^+$ and $W^-$ production rates are expected to be modified -- which is often referred to as the isospin effect -- but not their sum.
Furthermore, EW boson production is sensitive to the parton distribution functions~(PDF) which define the initial kinematics of the hard process.  In \PbPb\ collisions, production of $W^\pm$ bosons may differ from that in \pp\ collisions due to effects arising from the presence of the bound nucleons in the nucleus.
The measurements of $W^\pm$ boson production in heavy-ion collisions therefore offer an opportunity to extract valuable information about nuclear modifications to the free-nucleon PDF~\cite{Vogt:2000hp,Paukkunen:2010qg, CMS-HIN-13-007, Sirunyan:2019dox}.
 
Leptonic $W^\pm$ boson decays are of particular interest, since the charged leptons are expected to not interact substantially with the QGP.
Differences between the angular distributions of the $W^+$ and $W^-$ boson decay products and the different relative yields of $W^+$ and $W^-$ bosons produced in Pb+Pb and \pp\ collisions can be explored using lepton charge asymmetry. This observable is defined as the difference between the differential yields of positively and negatively charged leptons divided by their sum, expressed as a function of the charged-lepton pseudorapidity~($\eta_\ell$\footnote{
ATLAS uses a right-handed coordinate system with its origin at the nominal interaction point~(IP)
in the centre of the detector and the $z$-axis along the beam pipe. The $x$-axis points from the IP to the centre of the LHC ring,
and the $y$-axis points upward. Cylindrical coordinates $(r,\phi)$ are used in the transverse plane, $\phi$ being the azimuthal angle
around the $z$-axis. The pseudorapidity is defined in terms of the polar angle $\theta$ as $\eta=-\ln\tan(\theta/2)$.}):
\begin{equation}
\nonumber
A_\ell(\eta_\ell) = \frac{\der N_{W^+\to\ell^+\nu}/\der\eta_\ell - \der N_{W^-\to\ell^-\bar{\nu}}/\der\eta_\ell}{\der N_{W^+\to\ell^+\nu}/\der\eta_\ell + \der N_{W^-\to\ell^-\bar{\nu}}/\der\eta_\ell}.
\label{eq:Al}
\end{equation}
In the measurement of the asymmetry as a function of $\eta_\ell$, several systematic effects are reduced significantly in the ratio.

In a nucleus--nucleus~(A+A) collision in the absence of nuclear effects, the number of events of a hard process $X$ (\Nx) is proportional to the \pp\ cross-section for this process ($\sigma_{X}^{pp}$) scaled by factors related to the A+A collision geometry. These geometric parameters can be estimated using the Glauber approach~\cite{Alver:2008aq,Miller:2007ri} as detailed in Section~\ref{sec:hi-data}. An effective nucleon--nucleon (NN) cross-section for the process $X$ in A+A collisions, further referred to as normalised production yield, can be defined using the total number of inelastic A+A collisions (\Nevt) and the mean nuclear thickness function \avTaa (defined as the mean number of binary collisions divided by the total inelastic NN cross-section):
 
\begin{equation}
\sigmaNN=\frac{\Nx}{\Nevt\cdot\avTaa}.
\label{eq:sigmaNN}
\end{equation}
 
This expression allows a direct comparison between the production yields in heavy-ion collisions and the \pp\ cross-section for the same hard process. Differences between \sigmaNN\ and $\sigma_{X}^{pp}$ may arise due to nuclear effects including the isospin effect and inaccuracies in the description of the nuclear geometry. These differences are usually quantified using the nuclear modification factor defined as:
\begin{equation}
\Raa = \sigmaNN/\sigma_{X}^{pp}.
\label{eq:raa}
\end{equation}

In this paper, the first measurement of $W^\pm$ boson production yields in the electron and muon decay channels in \PbPb\ collisions at \hihienergy\ is presented. The data sample was collected in 2015 and corresponds to an integrated luminosity of $0.49 \pm 0.03~\nb^{-1}$. The normalised fiducial production yields are measured separately for $W^+$ and $W^-$ bosons. Then, the nuclear modification factors are extracted using production cross-sections in \pp\ collisions at the same centre-of-mass energy taken from Ref.~\cite{HION-2018-02-witherratum}.
The lepton charge asymmetry is measured as a function of the absolute value of the charged-lepton pseudorapidity. The results are compared with predictions based on next-to-leading-order~(NLO) QCD calculations with the use of  \textsc{CT14 NLO}~\cite{Dulat:2015mca} PDFs (accounting for the isospin effect) and two sets of predictions including nuclear modifications: EPPS16~\cite{Eskola:2016oht} and nCTEQ15~\cite{Kovarik:2015cma}.
 
This paper is organised as follows. The ATLAS detector is introduced in Section~\ref{sec:atlas}. The data and simulated event samples are described in Section~\ref{sec:hi-mc}. A brief discussion of the data analysis procedure and systematic uncertainties is given in Section~\ref{sec:data-analysis}. The results are presented in Section~\ref{sec:results} and the paper is summarised in Section~\ref{sec:summary}.
% End of text imported from the .//Sec_1.tex input file
% The next lines are included from the .//Sec_2.tex input file
\section{The ATLAS detector}
\label{sec:atlas}
The ATLAS detector~\cite{PERF-2007-01} at the LHC covers nearly the entire solid angle around the collision point.
It consists of an inner tracking detector surrounded by a thin superconducting solenoid, electromagnetic and hadronic calorimeters
and a muon spectrometer incorporating three large superconducting toroidal magnets.
 
The inner-detector system~(ID) is immersed in a \SI{2}{\tesla} axial magnetic field
and provides charged-particle tracking in the range $|\eta| < 2.5$.
A high-granularity silicon pixel detector including the insertable B-layer installed before Run~2~\cite{ATLAS-TDR-2010-19, PIX-2018-001}
covers the vertex region and is followed by a silicon microstrip tracker.
These silicon detectors are complemented by a transition radiation tracker,
which enables radially extended track reconstruction up to $|\eta| = 2.0$.
 
The calorimeter system covers the pseudorapidity range $|\eta| < 4.9$.
Within the region $|\eta|< 3.2$, electromagnetic~(EM) calorimetry is provided by
high-granularity lead/liquid-argon~(LAr) calorimeters,
with an additional thin LAr presampler covering $|\eta| < 1.8$
to correct for energy loss in material upstream of the calorimeters.
The EM calorimeter is divided into a barrel section covering $|\eta|<1.475$ and two
endcap sections covering $1.375<|\eta|<3.2$.
Hadronic calorimetry is provided by a steel/scintillator-tile calorimeter,
segmented into three barrel structures within $|\eta| < 1.7$ and two copper/LAr hadronic endcap calorimeters.
The solid-angle coverage is completed with  copper/LAr and tungsten/LAr forward calorimeter modules~(FCal)
in $3.1<|\eta|<4.9$, optimised for electromagnetic and hadronic measurements, respectively.
 
The muon spectrometer~(MS) comprises separate trigger and
high-precision tracking chambers measuring the deflection of muons in
the magnetic field generated by the toroidal magnets.
The field integral of the toroids ranges between \num{2.0} and \SI{6.0}{\tesla\metre}
across most of the detector.
A set of precision chambers covers the region $|\eta| < 2.7$ with three layers of monitored drift tubes,
complemented by cathode-strip chambers in the forward region.
The muon trigger system uses resistive-plate chambers in the barrel ($|\eta|<1.05$), and thin-gap chambers in the endcap ($1.05<|\eta|<2.4$) regions.
 
Two zero-degree calorimeters~(ZDC), situated at approximately $\pm140$ m from the nominal IP, detect neutral particles,
mostly neutrons and photons, with $|\eta|> 8.3$. The ZDC use tungsten plates as absorbers, and quartz rods sandwiched
between the tungsten plates as the active medium.
 
In 2015, the ATLAS detector had a two-level trigger system~\cite{TRIG-2016-01}. The level-1 trigger is implemented in
custom hardware and uses a subset of detector information to reduce the event
rate to a value of at most \SI{100}{\kHz}. This is followed by
a software-based high-level trigger
which further reduces the rate in order to record events to disk at about \SI{1}{\kHz}.
 
% End of text imported from the .//Sec_2.tex input file
% The next lines are included from the .//Sec_3.tex input file
\section{Data and simulated event samples}
\label{sec:hi-mc}
 
\subsection{Data sample and event centrality}
\label{sec:hi-data}
This analysis is based on the full set of \PbPb\ collision data collected by the ATLAS experiment in 2015 at a centre-of-mass energy of $\sqrtNN=5.02$~\TeV\ and corresponds to an integrated luminosity of 0.49~\invnb.

In heavy-ion measurements, centrality classes represent the percentiles of the total inelastic non-Coulombic cross-section excluding diffractive contributions~\cite{HION-2010-01}, and reflect the overlap volume of the two colliding nuclei.
In the ATLAS experiment, the centrality of \PbPb\ events is defined using the total transverse energy measured in the FCal, which is evaluated at the electromagnetic scale and denoted by FCal~\sumet. Geometric parameters, such as the average number of inelastically interacting nucleons in both colliding nuclei, \avNpart, and the average  nuclear thickness function, \avTaa, and their systematic uncertainties are obtained from the Glauber model~\cite{Miller:2007ri} and assigned to each centrality class by matching them to the modelled FCal~\sumet\ distribution.
 
The centrality determination for the 2015 \PbPb\ dataset follows a procedure similar to that used for lower-energy \PbPb\ data in ATLAS, which is described in Ref.~\cite{HION-2016-07}. In that procedure, the set of \PbPb\ collision geometries is defined via the Glauber model, using the Monte Carlo Glauber (MCGlauber) code v2.4~\cite{Loizides:2014vua}, an inelastic nucleon--nucleon cross-section of $\sigmaTotNN = 70\pm5$~mb at \hihienergy, and a single Woods--Saxon distribution for the nucleon radial profile. The modelled FCal~\sumet\ distribution is matched with the distribution measured in minimum-bias~(MB) data selected with FCal~$\sumet > 40$~\GeV. This FCal~\sumet\ selection ensures that contributions from photonuclear and diffractive events in the fit range are negligible. Table~\ref{tab:centrality} lists the values of the \avNpart\ and \avTaa\ parameters with their uncertainties in different centrality classes used in this analysis.

\begin{table}[hbt!]
\begin{center}
\caption{Geometric parameters extracted from the MCGlauber code v2.4 for different centrality classes in 2015 \PbPb\ data. Average numbers of inelastically interacting nucleons \avNpart\ and mean values of the nuclear thickness function \avTaa\ are listed with their absolute and relative uncertainties.}
\begin{tabular}{r|rl|rl}
Centrality & $\avNpart$ & $\delta\avNpart$  & $\avTaa$ [1/mb] & $\delta\avTaa$ \\
\hline
0--2\% & 399.0 $\pm$ 1.2 & 0.30\% & 28.30 $\pm$ 0.25 & 0.88\% \\
2--4\% & 380.2 $\pm$ 2.0 & 0.53\% & 25.47 $\pm$ 0.21 & 0.82\% \\
4--6\% & 358.9 $\pm$ 2.4 & 0.67\% & 23.07 $\pm$ 0.21 & 0.91\% \\
6--8\% & 338.1 $\pm$ 2.7 & 0.80\% & 20.93 $\pm$ 0.20 & 0.96\% \\
8--10\% & 317.8 $\pm$ 2.9 & 0.91\% & 18.99 $\pm$ 0.19 & 1.0\% \\
10--15\% & 285.2 $\pm$ 2.9 & 1.0\% & 16.08 $\pm$ 0.18 & 1.1\% \\
15--20\% & 242.9 $\pm$ 2.9 & 1.2\% & 12.59 $\pm$ 0.18 & 1.4\% \\
20--25\% & 205.6 $\pm$ 2.9 & 1.4\% & 9.77 $\pm$ 0.18 & 1.8\% \\
25--30\% & 172.8 $\pm$ 2.8 & 1.6\% & 7.50 $\pm$ 0.17 & 2.3\% \\
30--40\% & 131.4 $\pm$ 2.6 & 2.0\% & 4.95 $\pm$ 0.15 & 3.0\% \\
40--50\% & 87.0 $\pm$ 2.4 & 2.8\% & 2.63 $\pm$ 0.11 & 4.2\% \\
50--60\% & 53.9 $\pm$ 2.0 & 3.7\% & 1.28 $\pm$ 0.07 & 5.8\% \\
60--80\% & 23.0 $\pm$ 1.3 & 5.7\% & 0.39 $\pm$ 0.03 & 8.2\% \\
\hline
0--80\% 	& 141.3 $\pm$ 2.1 & 1.5\%	& 7.00 $\pm$ 0.11 	& 1.6\% \\
\end{tabular}
\label{tab:centrality}
\end{center}
\end{table}
 
For consistency with other heavy-ion measurements from the ATLAS Collaboration, this analysis uses the binning in FCal \sumet\ and geometric parameters determined from the MCGlauber code v2.4 as the default. Recently, an updated version of the MCGlauber code, v3.2, became available with several suggested improvements in the geometric modelling. These improvements are described in Ref.~\cite{Loizides:2017ack} and include a lower value of \sigmaTotNN\ at \hihienergy with a smaller uncertainty ($67.6\pm0.5$~mb), separate radial distributions for protons and neutrons in the nucleus, and other improvements in the determination of nucleon positions within the nucleus.
 
To reassess the scaling of boson yields within this improved model of the \PbPb\ collision geometry, the centrality determination is performed following the same procedure as described in Ref.~\cite{HION-2016-07} but using an alternative set of \PbPb\ events generated with the MCGlauber code v3.2. The fit to the FCal~\sumet\ distribution in data results in a larger estimate (by~0.6\%) for the fraction of inelastic \PbPb\ events being contained in MB events selected with FCal~$\sumet > 40$~\GeV. Because of the increased fraction of inelastic \PbPb\ events in the FCal~\sumet\ range used for the fit, each centrality range is mapped to a systematically higher range of FCal~\sumet\ values, and the number of MB \PbPb\ events in each centrality selection is lower by approximately 0.6\%. Furthermore, the estimated value of \avTaa\ in the centrality classes is lower by 1\% in the most central events but higher by 6--7\% in the most peripheral classes, consistent with the change in \avTaa\ found in Ref.~\cite{Loizides:2017ack}. The systematic uncertainties of \avTaa\ are determined following procedures identical to the MCGlauber code v2.4 case~\cite{HION-2016-07}, but with a smaller \sigmaTotNN\ variation of $\pm0.5$~mb. The binning in FCal \sumet\ and geometric parameters determined from the MCGlauber code v3.2 are used only for comparison with the MCGlauber code v2.4.

\subsection{Monte Carlo simulation of signal and background events}
 
Samples of Monte Carlo~(MC) simulated events were used to evaluate the selection efficiencies and to model the properties of signal and background processes. The response of the ATLAS detector was simulated using the \textsc{Geant4} framework~\cite{geant4, SOFT-2010-01}. Signal processes, i.e. the $W^\pm$ boson production and leptonic decays, were modelled with the \POWHEGBOX\ v2 event generator~\cite{Alioli:2010xd, Alioli:2008gx} interfaced to \PYTHIA\ 8.186~\cite{Sjostrand:2007gs} to model parton showering and fragmentation processes. The CT10 PDF set~\cite{Lai:2010vv} evaluated with NLO accuracy was used to set the initial kinematics for the matrix-element calculation. Events produced in EW background processes ($W^\pm\to\tau^\pm\nu$, $Z\to\mu^+\mu^-$, $Z\to e^+e^-$, $Z\to\tau^+\tau^-$) were generated with the same generator set-up. To model the production of top-quark pairs~($t\bar{t}$), the \POWHEGBOX\ v2 event generator~\cite{Frixione:2007nw} was used with the CT10f4 PDF set~\cite{Lai:2010vv} in the matrix element. For $t\bar{t}$ production, the parton shower and fragmentation were simulated using \PYTHIA\ 6.428~\cite{Sjostrand:2006za-correct}. More details of the event generator set-up for all considered processes can be found in Ref.~\cite{HION-2018-02-witherratum}, where the same configurations were used.
 
MC subsamples were produced separately for $pp$, $pn$, $np$, and $nn$ collisions and a weighting procedure was applied to combine all subsamples as described here. For each subprocess of interest, a global event weight was derived: it is based on the mass~($A=208$) and atomic~($Z=82$) numbers of the colliding lead nuclei, and on the total number of generated events. This corresponds to a fraction of all nucleon-nucleon collisions of $f_{pp}=(Z/A)^2=15.5\%$ for $pp$, $f_{pn} = f_{np}=Z(A-Z)/A^2=23.9\%$ for $pn$ or $np$, and $f_{nn}=[(A-Z)/A]^2=36.7\%$ for $nn$. The global event weight for each subprocess is calculated as the ratio of the number of expected events to the number of generated events:
\begin{equation}
\nonumber
w = \frac{\avgTAA^{0-80\%}\cdot N^{0-80\%}_{\mathrm{evt, MB}}\cdot\sigma_{ij}}{N^{0-80\%}_{\mathrm{gen}, ij}} f_{ij},
\label{eq:isospin}
\end{equation}
\noindent
where $f_{ij}$ stands for $f_{pp}$, $f_{pn,np}$ or $f_{nn}$, $N^{0-80\%}_{\mathrm{gen}, ij}$ is the number of generated events for the given subprocess in the 0--80\% centrality class~(see Section~\ref{sec:hi-data}), $N^{0-80\%}_{\mathrm{evt, MB}}$ is the total number of MB events in the 0--80\% centrality class, and $\sigma_{ij}$ is the production cross-section for the given subprocess. If not stated otherwise, theoretical predictions presented in this paper are calculated using fixed fractions of $pp$, $pn$, $np$, and $nn$ collisions.
For the background subtraction procedure, $W^\pm$ and $Z$ boson production cross-sections are scaled to NNLO accuracy using DYNNLO~\cite{Catani:2007vq,Catani:2009sm} calculations with the \textsc{CT14 NNLO} PDF set~\cite{Dulat:2015mca}. The scaling factors take the following values: 1.026 for $W^+$, 1.046 for $W^-$, and 1.007 for $Z$ boson production. It should be noted that the DYNNLO code supports calculations only for the \pp\ isospin combination. For other isospin combinations, the same scaling as for \pp\ collisions is assumed.
 
In order to study detector performance in conditions that match the data, the simulated events were embedded into experimental data taken during the 2015 \PbPb\ run. This data-overlay procedure ensures an accurate description of the underlying event in the MC simulation, and additionally provides detector conditions matching those of Pb+Pb data-taking periods. Events used in the overlay procedure were recorded using a dedicated set of MB triggers and total~\sumet\ triggers, which were used to enhance the rate of more central events.
% End of text imported from the .//Sec_3.tex input file
% The next lines are included from the .//Sec_4.tex input file
\section{Data analysis}
\label{sec:data-analysis}
 
\subsection{Object definitions and event selection}
\label{sec:event-selection}
Candidate events with $W^\pm$ boson production are required to have only one primary vertex reconstructed from at least three tracks with a transverse momentum, \pt, larger than 400~\MeV, and to pass a trigger selection, which requires a single electron or muon candidate with a \pt threshold of 15~\GeV\ or 8~\GeV, respectively. In addition,  the electron trigger applies a loose identification requirement~\cite{PERF-2017-01_corr} and the underlying-event contribution to the energy deposits in calorimeter cells is subtracted~\cite{HION-2011-02}.

In events assigned to the 50--80\% centrality classes, an additional selection is made using the ZDC in order to suppress EM background contributions. Events with at least one neutron detected in each arm of the ZDC are accepted, and the fraction of rejected events is about 0.4\% in both the electron and muon channels.
 
A small fraction of the selected events contain more than one inelastic interaction (pile-up). The anti-correlation between the FCal~\sumet\ and the number of neutrons detected in the ZDC is used to suppress pile-up events. Events with a number of ZDC neutrons much higher than the number expected from the bulk of events for a given value of FCal~\sumet\ are rejected. The fraction of rejected events is about 0.4\% in both the electron and muon channels, and is constant across centrality classes.

The electron trigger efficiency is 99\% for peripheral events and slowly decreases to 96\% for central events. The muon trigger efficiency in the endcap region of the detector is $\sim$90\% and in the barrel region it varies from 60\% to 80\%. No dependence on detector occupancy (centrality) is found.
 
Electron candidates are reconstructed using information from tracking detectors and the EM calorimeter~\cite{PERF-2017-01_corr}. They are required to have $\pt^\ell>25$~\GeV\ and $|\eta_\ell|<2.47$. Candidates falling in the transition region between barrel and endcap calorimeters ($1.37<|\eta_\ell|<1.52$) are rejected. In addition, isolation and `medium' likelihood-based identification requirements~\cite{PERF-2017-01_corr} optimised for \PbPb\ collisions as a function of centrality are applied. Muon candidates are reconstructed by combining tracks measured in the ID with tracks measured in the MS~\cite{PERF-2015-10}, and must satisfy $\pt^\ell>25$~\GeV\ and $|\eta_\ell|<2.4$. In addition, muons have to pass the requirements of `medium' identification and of a dedicated isolation selection~\cite{PERF-2015-10}.

The electron energy calibration is primarily obtained from the simulation by employing multivariate techniques~\cite{PERF-2017-03}. Residual corrections to the energy scale and resolution are determined from data by comparing the measured \zee\ invariant mass distribution to the one predicted by the simulation~\cite{PERF-2017-03}. This procedure was found to be sensitive to the pile-up distribution in data due to different settings used for the signal readout from the EM calorimeters~\cite{Abreu:2010zzc}. Therefore, a special set of energy scale correction factors was derived for the 2015 Pb+Pb dataset. Measurements of muon momenta can be biased by the detector alignment and resolution, distortions of the magnetic field or imprecise estimates of the amount of passive material in the detector. Corrections of the  muon momentum scale and resolution, which are applied to the simulation, are derived using \zmumu events~\cite{PERF-2015-10}.
 
Events with $W^\pm$ boson candidates are selected by requiring an electron or a muon that is matched to a lepton
selected at the trigger level. The (anti-)neutrinos from $W^\pm \to \ell^\pm\nu$ decays escape direct detection. A measure of the neutrino transverse momentum, \ptnu, can be inferred from the global event momentum imbalance in the plane transverse to the beam axis.
In heavy-ion collisions, low-\pt\ particle production is significantly enhanced compared to \pp\ collisions, thereby resulting in a resolution of missing transverse momentum obtained from calorimeter cells that is much worse than in \pp\ collisions. In the most central \PbPb\ collisions at $\sqrtNN = 2.76$~\TeV, the resolution reaches as much as 45~\GeV~\cite{HION-2013-05} due to enhanced contributions from the underlying event, while for $\sqrtNN = 5.02$~\TeV, it is expected to be even larger because of  increased underlying-event activity. Therefore, tracks are used instead of calorimeter cells, as low-\pt\ tracks from the underlying event can be suppressed more easily. The missing transverse momentum vector, \ptmiss, is defined as the negative vector sum of the ID-track transverse momenta, excluding good leptons with a poor-quality ID track. In the case of electrons, the calorimeter energy measurement is used, while for muons the \pt\ determined from a combined reconstruction using ID and MS hits is used. This approach is analogous to the one developed in \pp\ collisions~\cite{PERF-2016-07}. In order to minimise the noise contribution from the underlying event while
retaining sensitivity to the contribution from the hard-scattering process, only tracks with $\pt>4$~\GeV\ are used in the calculation of \ptmiss.
The transverse mass of the lepton$-\ptmiss$ system is defined as:
\begin{equation}
\nonumber
\mt = \sqrt{2\pt^\ell\ptmiss\left(1-\cos{\Delta\phi}\right)},
\label{eq:mT}
\end{equation}
\noindent
where $\Delta\phi$ is the azimuthal angle between $\pt^\ell$ and \ptmiss vectors. The $W^\pm$ boson candidates are required to have $\ptmiss>25$~\GeV\ and $\mt>40$~\GeV.
 
The background contribution from $Z\to\ell^+\ell^-$ decays is further suppressed by imposing a $Z$-veto requirement. Events with at least two leptons of the same flavour which form an opposite-charge pair with an invariant mass above 66~\GeV\ are rejected. These events are selected by requiring that one lepton in the pair has $\pt>25$~\GeV\ and fulfils all other quality criteria discussed above, while the other lepton in the pair passes a lower \pt\ threshold of 20~\GeV\ with looser quality requirements.

\subsection{Background estimation}
\label{sec:bkgr}
Background processes that contribute to the $W^\pm$ boson production measurement are EW processes producing $W^\pm\to\tau^\pm\nu$, $Z\to\ell^+\ell^-$ and $Z\to\tau^+\tau^-$ decays, as well as top-quark production and multi-jet processes. The multi-jet background includes various processes such as semileptonic decays of heavy-flavour hadrons or in-flight decays of kaons and pions for the muon channel, and photon conversions or misidentified hadrons for the electron channel.
 
The background contributions from EW production are evaluated using simulated event samples described in Section~\ref{sec:hi-mc}. They are normalised according to their expected number of events in the data evaluated from Eq.~\eqref{eq:sigmaNN} using production cross-sections scaled to NNLO accuracy. It is found that the contributions from $Z\to\ell^+\ell^-$ and $W^\pm\to\tau^\pm\nu$ processes dominate. In the electron channel, they amount to 4.1\% and 1.6\%, respectively, for electrons, while for positrons, these fractions are 4.2\% and 1.5\%, respectively. In the muon channel, they amount to 3.0\% and 1.9\%, respectively, of the event sample selected with negative muons, while for positive muons, these fractions are 3.1\% and 1.8\%, respectively.
 
The background contributions from $t\bar{t}$ production are also evaluated using MC simulation. They are estimated to be at the level of 0.1\% for electrons and 0.2\% for muons. Contributions from the production of single top quarks and dibosons, which are even smaller, are neglected.
 
A large fraction of multi-jet background events are rejected by the lepton isolation requirement and the \ptmiss\ selection. However, the very large production cross-sections for multi-jet processes make their contribution to the selected event sample significant. This contribution is estimated using template fits to $\pt^\ell$ distributions for $\pt^\ell > 20$~\GeV\ following a method similar to the one described in Refs.~\cite{STDM-2014-18, HION-2018-02-witherratum}.
Template distributions enriched in events from multi-jet background processes are taken from data by selecting events with non-isolated leptons, while templates for the signal and other background processes are extracted from the MC simulation.
The variable used to determine the isolation of electrons and muons is, however, correlated with $\pt^\ell$ for multi-jet events, modifying the $\pt^\ell$ distribution shape for non-isolated leptons relative to that for isolated leptons. Therefore, prior to the fit, the shape of the multi-jet background template in $\pt^\ell$ is corrected, so that it more closely matches the shape of the multi-jet background distribution passing the signal isolation selection. The correction procedure is given below for the muon channel. In the case of the electron channel, all steps of the procedure are similar.
 
\begin{figure}[tbh!]
\centering
\includegraphics[width=0.49\textwidth]{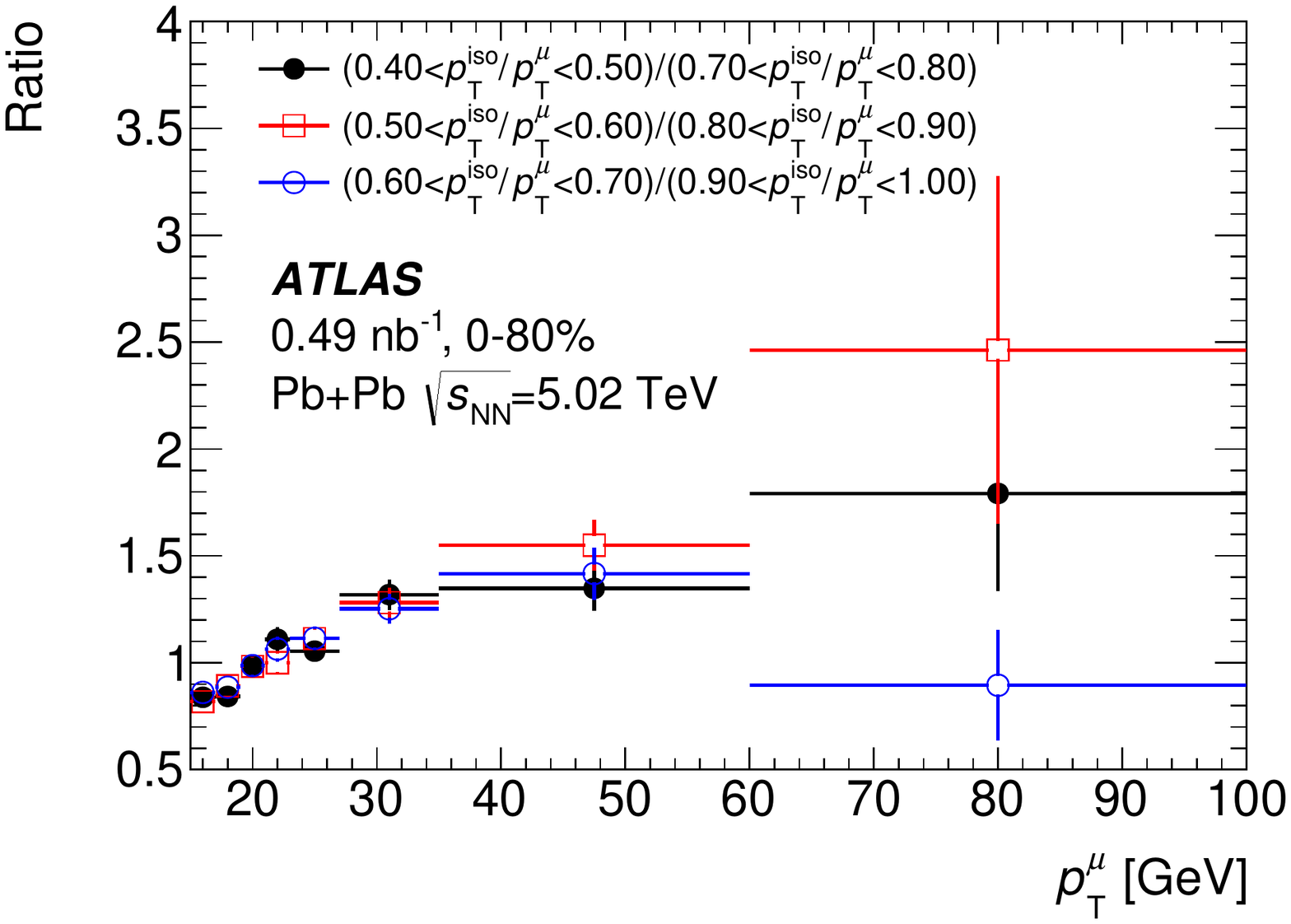}
\caption{Ratios of the $\pt^\mu$ distribution shape for multi-jet background templates extracted from different ranges of $\pt^\mathrm{iso}/\pt^\mu$. The ratios are taken between templates from ranges of $\pt^\mathrm{iso}/\pt^\mu$ which have centres separated by 0.3 units. The error bars represent the statistical uncertainties.}
\label{fig:QCD-MJ-weight-muons}
\end{figure}

The events with non-isolated muons are divided into subsamples defined by ranges  of 0.1 unit in a track-based isolation variable, $\pt^\mathrm{iso}/\pt^\mu$, where $\pt^\mathrm{iso}$ is the sum of track transverse momenta in a cone around the muon. From each of the subsamples, a multi-jet background template is extracted. The evolution of the template shapes is summarised in Figure~\ref{fig:QCD-MJ-weight-muons}, which shows ratios of self-normalised $\pt^\mu$ distributions for different templates. The ratios are taken between templates from ranges of $\pt^\mathrm{iso}/\pt^\mu$ which have centres separated by 0.3 units. The average of the ratios, $r$, is then used as a weight to correct the distribution shape of the multi-jet background:
\begin{equation}
\nonumber
N_\mathrm{MJ}^\mathrm{template}\left(\pt^\mu\right) = N_\mathrm{MJ}^\mathrm{uncorrected}\left(\pt^\mu\right) \cdot r\left(\pt^\mu\right)^{d/0.3}.
\end{equation}
\noindent
The weight $r$ is modified using the ratio of the distance $d$ between the centre of a given  $\pt^\mathrm{iso}/\pt^\mu$ range and the mean value of the signal isolation (determined from MC simulation) to the distance of 0.3 units between centres of  $\pt^\mathrm{iso}/\pt^\mu$ ranges used to determine the weight.
This procedure ensures that the extracted multi-jet background yields are stable regardless of the exact definition of the non-isolated muons used to construct a template. In order to estimate the multi-jet background yield differentially in $\eta_{\mu}$, the template fits are performed separately for each $\eta_{\mu}$ bin.
 
Events from multi-jet background processes are estimated to contribute up to about 20\% and 12\% to the event samples selected in the electron and muon channels, respectively. The multi-jet background fraction increases by about 10\% of the total in central collisions compared with peripheral collisions in both electron and muon channels.
 
\begin{figure}[tbh!]
\centering
\includegraphics[width=0.49\textwidth]{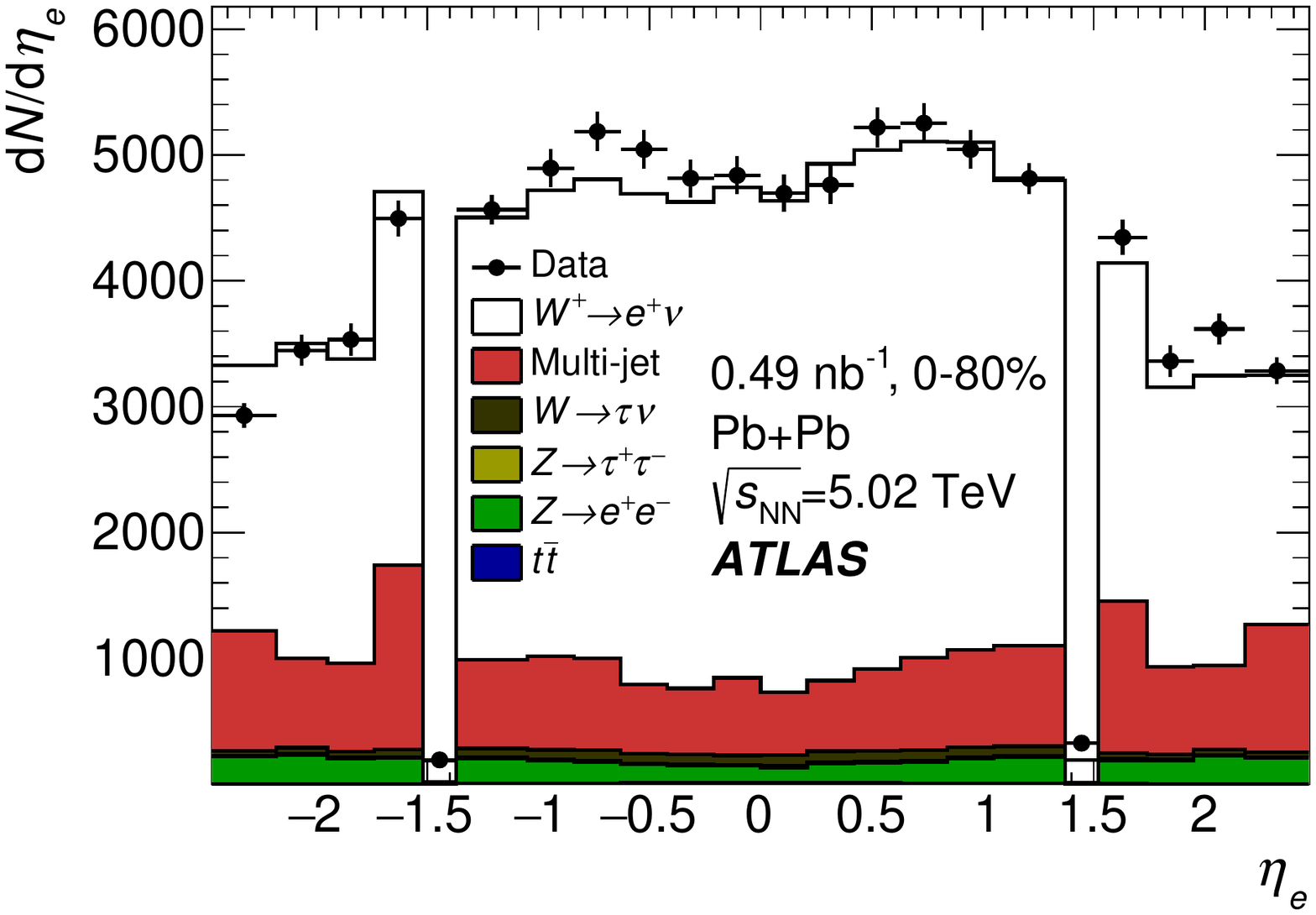}
\hfill
\includegraphics[width=0.49\textwidth]{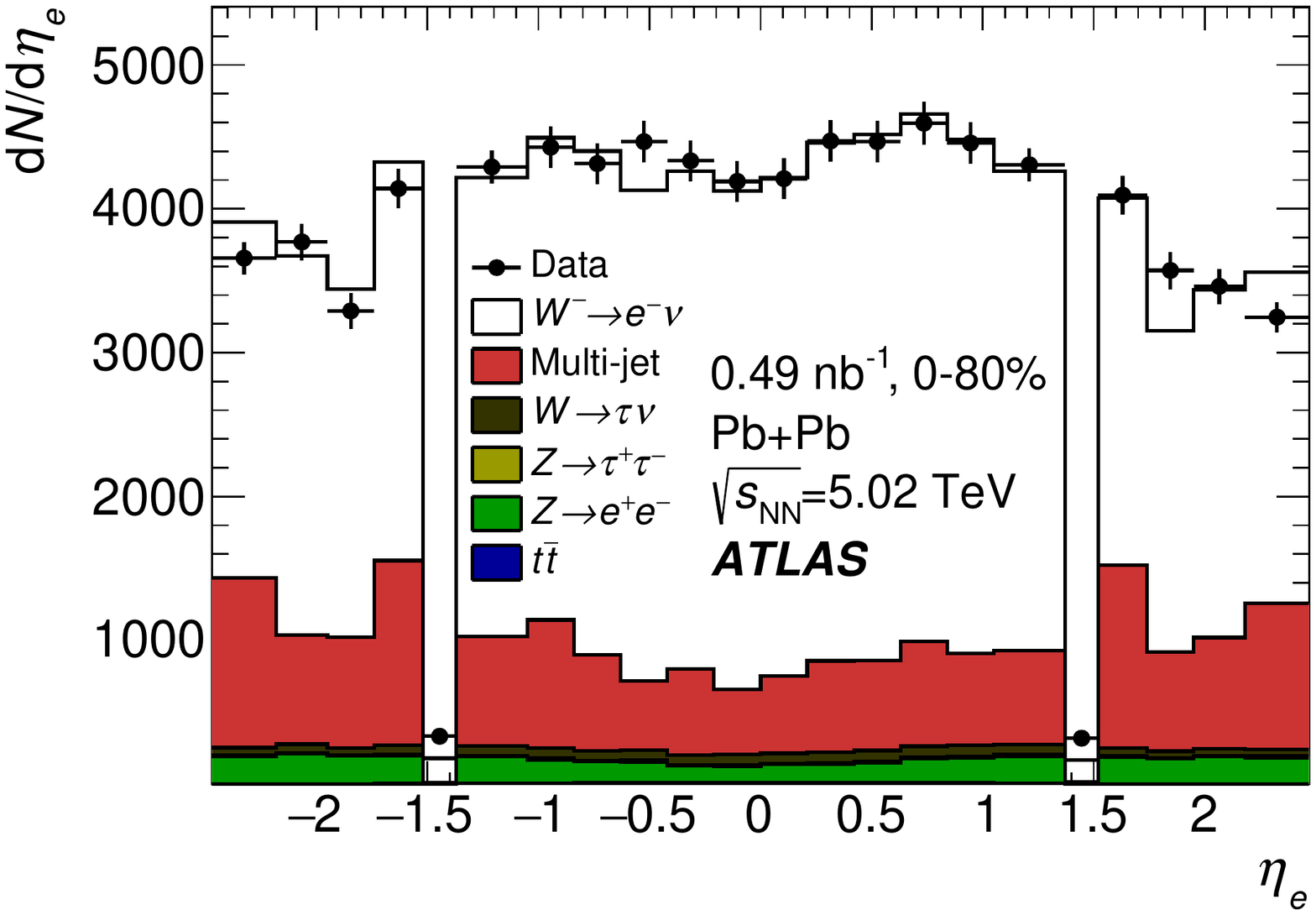}
\newline
\includegraphics[width=0.49\textwidth]{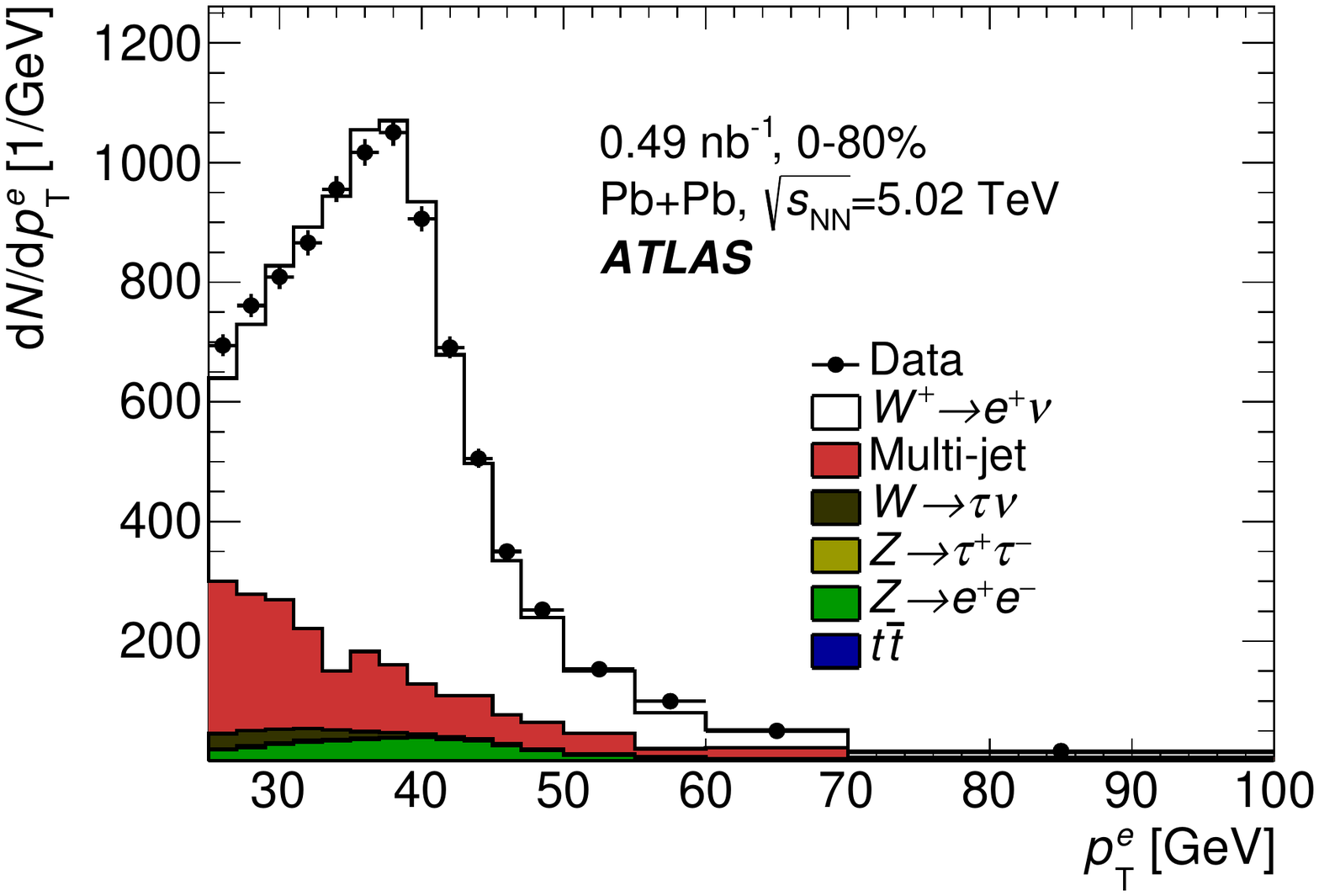}
\hfill
\includegraphics[width=0.49\textwidth]{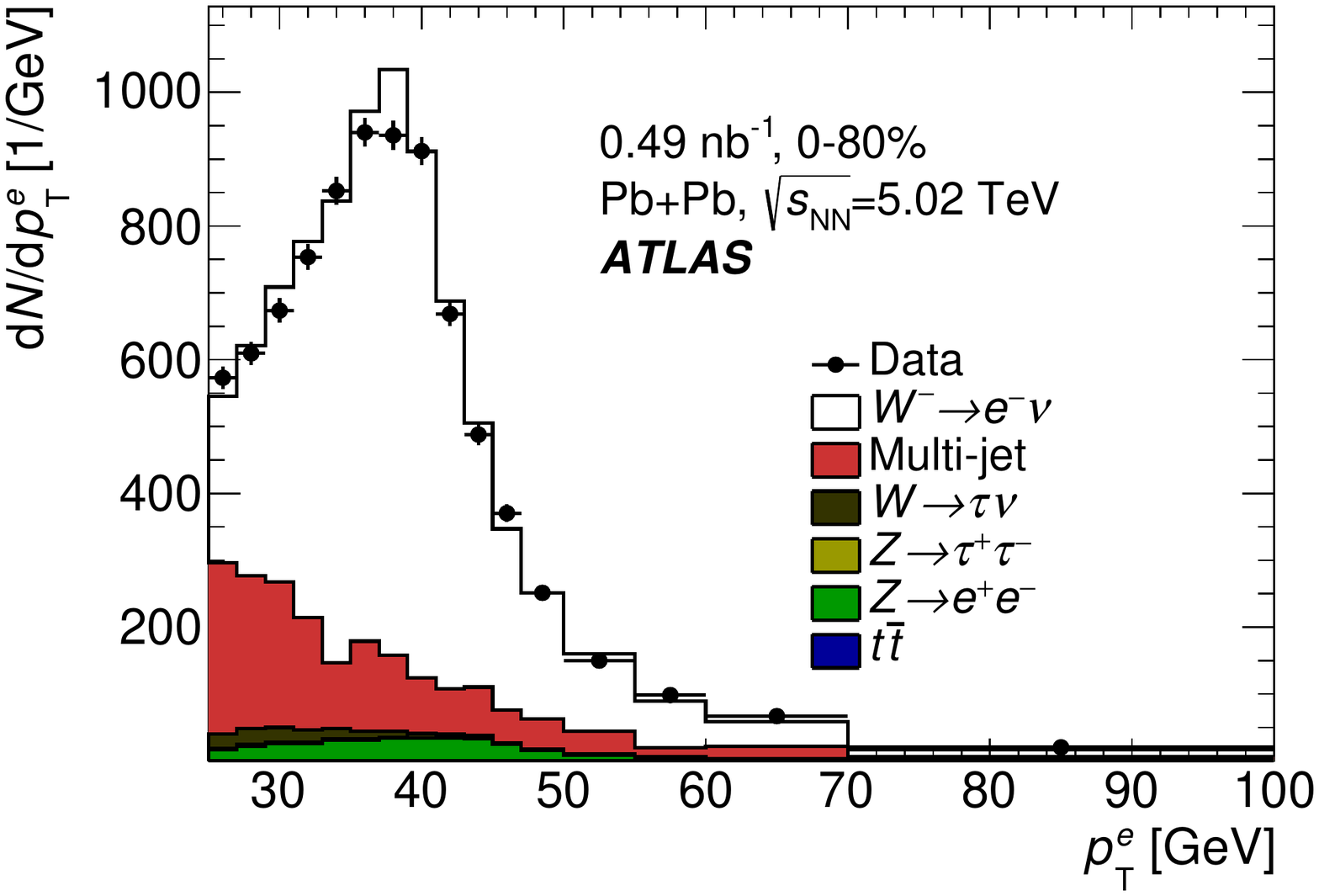}
\caption{Detector-level distributions of $W^{+}$~(left) and $W^{-}$~(right) event candidates decaying in the electron channels after all selection requirements as a function of the electron pseudorapidity~(top) and transverse momentum~(bottom). The contributions of EW and top-quark backgrounds are normalised according to their expected number of events in the data, while the contribution of QCD multi-jet background is normalised using a template fit to the $\pt^e$ distribution. Distributions are presented for the 0--80\% centrality class. The error bars represent the statistical uncertainties.}
\label{fig:control-plots-ele}
\end{figure}

\begin{figure}[thb!]
\centering
\includegraphics[width=0.49\textwidth]{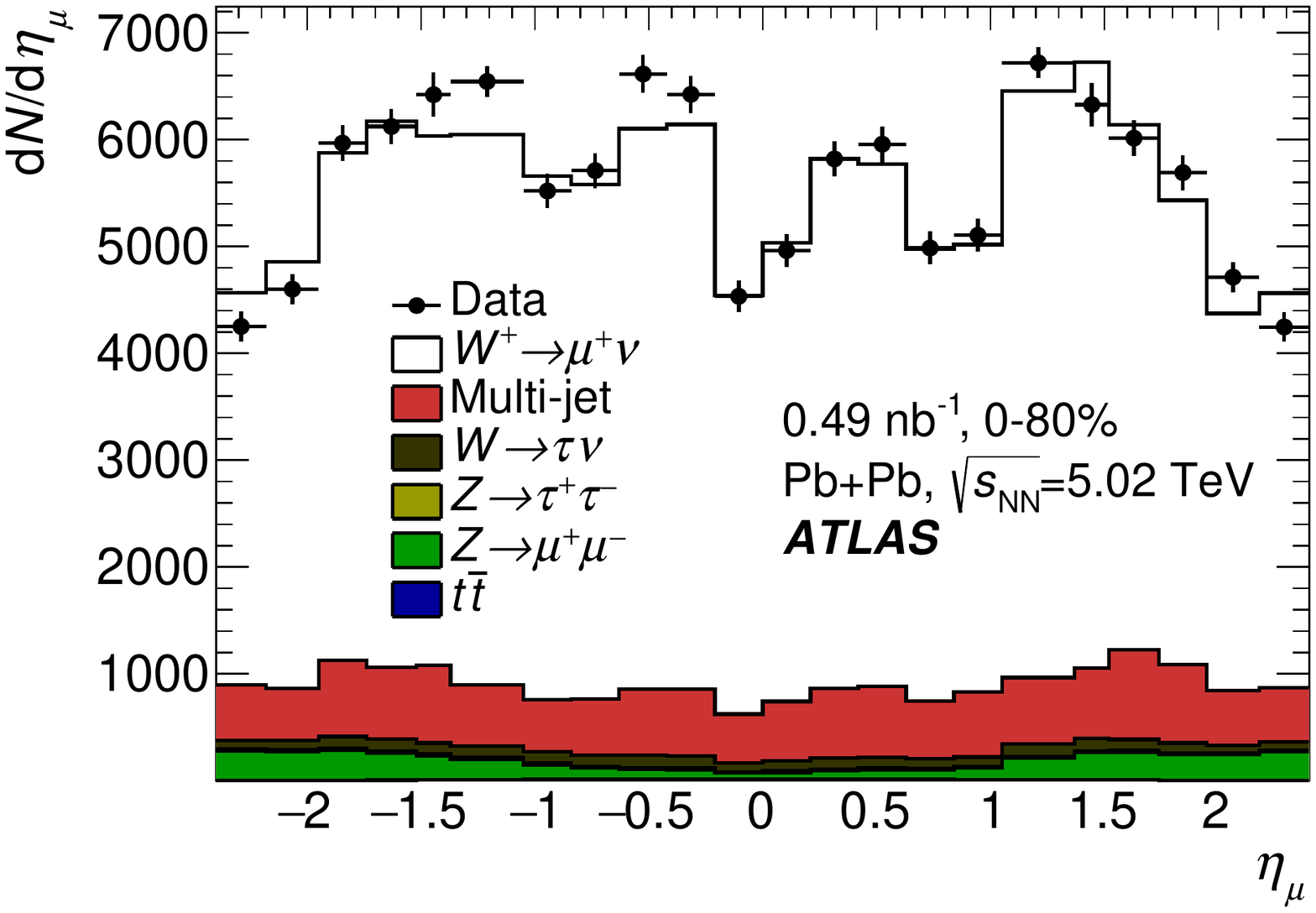}
\hfill
\includegraphics[width=0.49\textwidth]{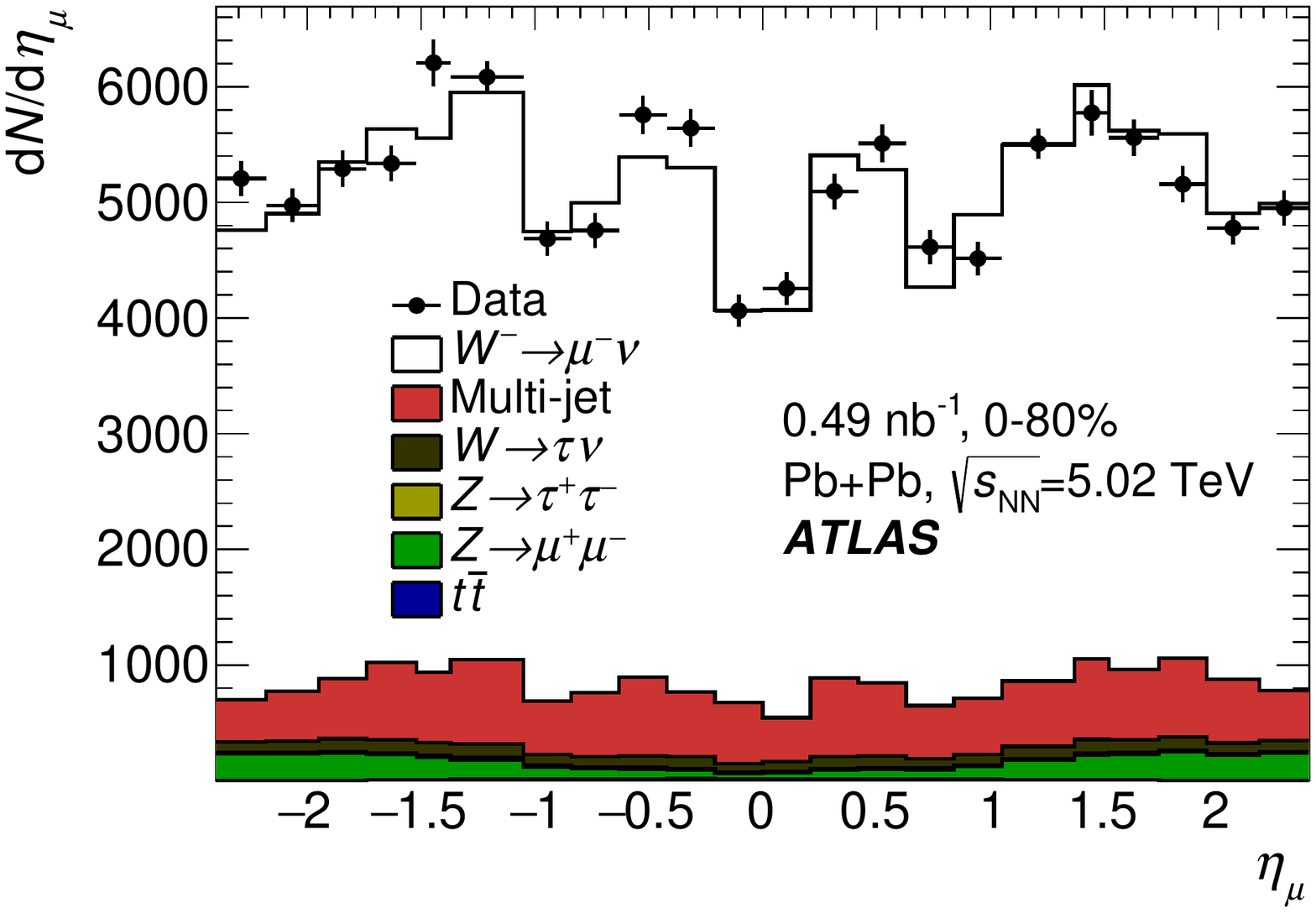}
\newline
\includegraphics[width=0.49\textwidth]{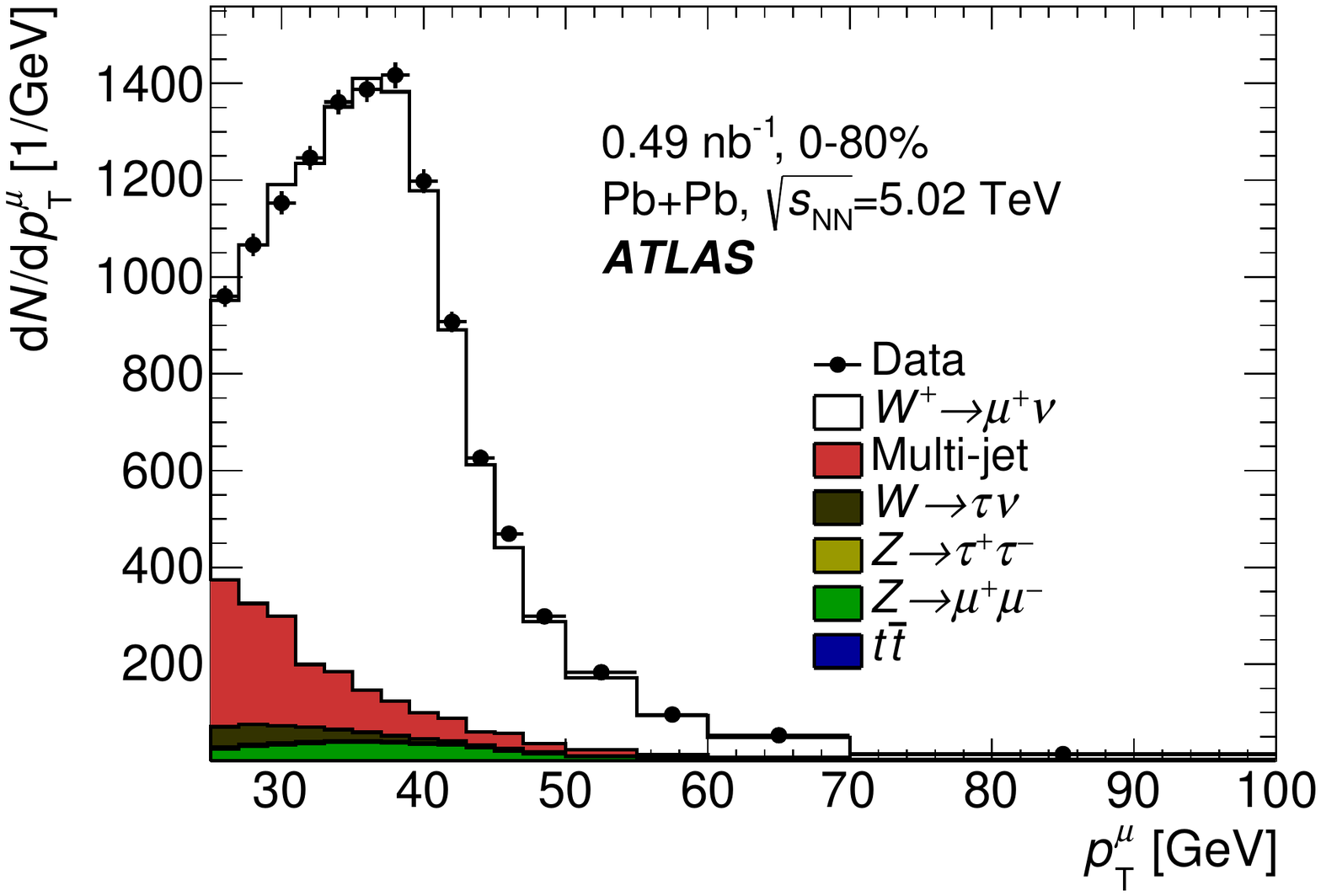}
\hfill
\includegraphics[width=0.49\textwidth]{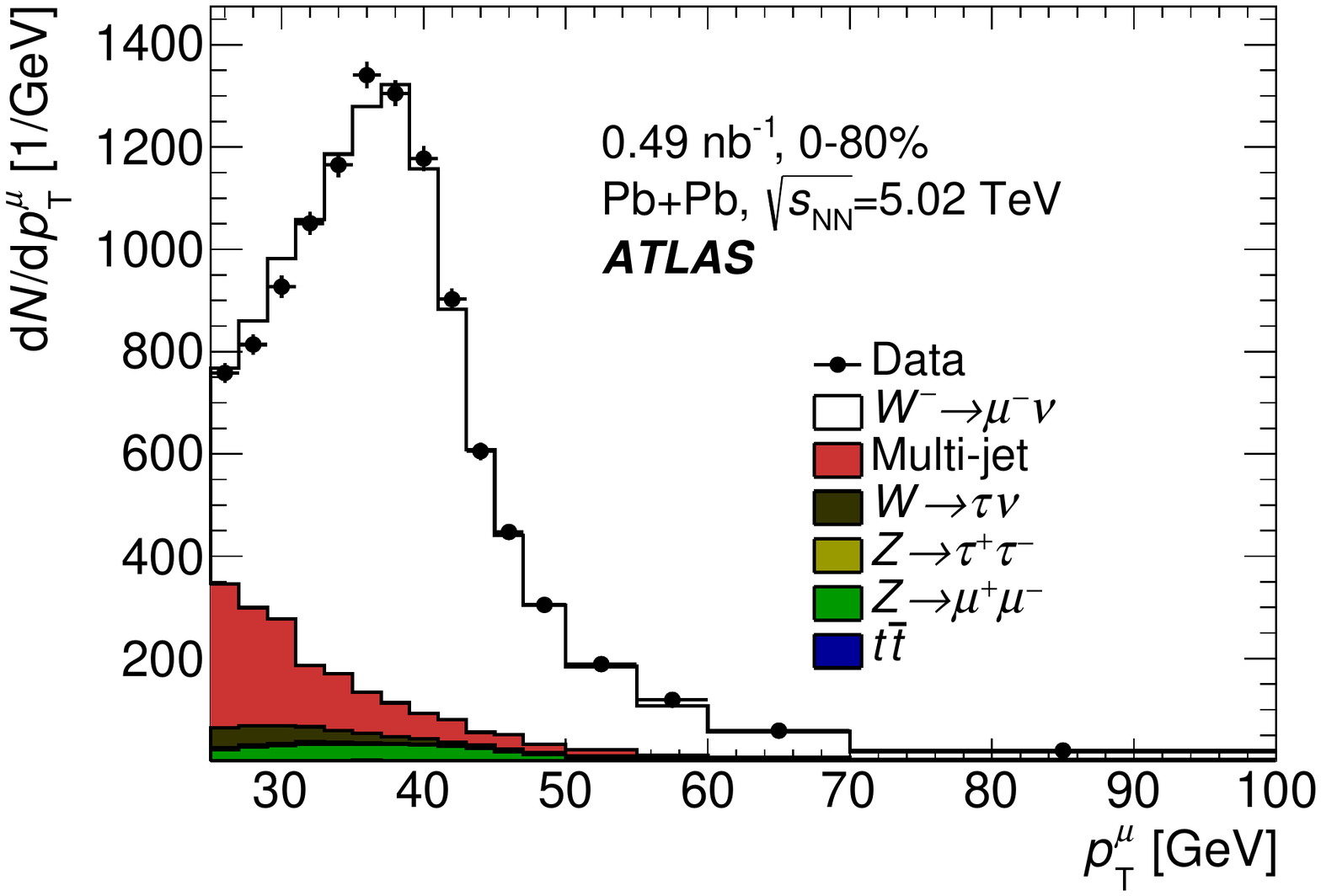}
\caption{Detector-level distributions of $W^{+}$~(left) and $W^{-}$~(right) event candidates decaying in the muon channels after all selection requirements as a function of the muon pseudorapidity~(top) and transverse momentum~(bottom). The contributions of EW and top-quark backgrounds are normalised according to their expected number of events in the data, while the contribution of QCD multi-jet background is normalised using a template fit to the $\pt^\mu$ distribution. Distributions are presented for the 0--80\% centrality class. The error bars represent the statistical uncertainties.}
\label{fig:control-plots-mu}
\end{figure}

Figures~\ref{fig:control-plots-ele} and \ref{fig:control-plots-mu} show the detector-level distributions of $W^{+}$ and $W^{-}$ event candidates decaying in the electron and muon channels, respectively, as a function of lepton pseudorapidity, $\eta_\ell$, and as a function of lepton transverse momentum, $\pt^\ell$. Background contributions from QCD multi-jet production and from EW processes discussed above are also shown in the plots. Fairly good agreement is found between data and the sum of signal and background contributions. The non-smooth behaviour of the multi-jet background distributions is related to large statistical uncertainties of the templates, which are propagated through the background subtraction procedure to the final results.
 
\FloatBarrier
\subsection{Experimental corrections}
\label{sec:corrections}
 
The \wlnu\ production yields in the electron and muon decay channels are measured in a fiducial phase-space region defined as:
\begin{equation}
\nonumber
\pt^\ell>25\,\mbox{\GeV},\;\;\;\;\;\;\;\; |\eta_\ell|<2.5,\;\;\;\;\;\;\;\; \ptnu>25\,\mbox{\GeV},\;\;\;\;\;\;\;\; \mt>40\,\mbox{\GeV},
\end{equation}
\noindent
where $\ell=e,\mu$ stands for the electron or muon, $\ptnu$ is the transverse momentum of the respective (anti)neutrino and $\mt$ is the transverse mass of the lepton and neutrino system. To correct for QED final-state emissions, the lepton kinematics are evaluated before photon radiation.
 
The \wlnu\ event yields are extracted in each bin of $\eta_\ell$ and centrality using the formula:
\begin{equation}
\nonumber
N_{W} = \frac{N_{W}^\mathrm{obs} - N_{W}^\mathrm{bkg}}{\Cw},
\end{equation}
\noindent
where $N_{W}^\mathrm{obs}$ and $N_{W}^\mathrm{bkg}$ are the numbers of observed and background events, respectively, and \Cw\ denotes bin-by-bin correction factors, which are evaluated using the signal MC simulation in bins of $\eta_\ell$ and centrality, accounting for differences between data and MC simulation as described below.  The correction factors are determined separately for each lepton charge and each decay channel, and are defined as:
\begin{equation}
\nonumber
\Cw\left(\eta_\ell^\mathrm{reco},\mathrm{centrality}\right)=\frac{N_{W}^{\mathrm{sel},\mathrm{pass}}\left(\eta_\ell^\mathrm{reco},\mathrm{centrality}\right)}{N_{W}^{\mathrm{sel},\mathrm{gen}}\left(\eta_\ell^\mathrm{true},\mathrm{centrality}\right)},
\end{equation}
\noindent
with $N_{W}^{\mathrm{sel},\mathrm{pass}}$ being the sum of event weights for events that fulfil the detector-level selection criteria described in Section~\ref{sec:event-selection}, while $N_{W}^{\mathrm{sel},\mathrm{gen}}$ denotes the sum of event weights for events selected in the generator-level fiducial phase space.
The \Cw\ correction factors account for differences between selections applied to the reconstructed lepton pseudorapidity, $\eta_\ell^\mathrm{reco}$, and the true pseudorapidity, $\eta_\ell^\mathrm{true}$.
These factors account also for the lepton reconstruction, identification, isolation, and trigger efficiencies, which are evaluated separately, as well as for the \ptmiss\ selection efficiency. Lepton efficiencies are measured in the data and determined in MC simulation using the tag-and-probe method in $Z\to\ell^+\ell^-$ events in the \PbPb\ system~\cite{PERF-2017-01_corr, PERF-2015-10, TRIG-2016-01}. They are evaluated as a function of the reconstructed $\eta_\ell$ and $\pt^\ell$ in the electron channel, while in the muon channel, they depend only on $\eta_\ell$. Differences between efficiencies extracted from the data and MC simulation do not exceed a few percent. Scale factors used to correct the MC simulation are derived as ratios of efficiencies determined in data and simulation. Within the precision of the tag-and-probe method, no dependence of scale factors on centrality is observed. The reconstructed sum of event weights, $N_{W}^{\mathrm{sel},\mathrm{pass}}$, is evaluated after correcting the simulation, such that the simulated detector response matches the response observed in data.

\begin{figure}[!b]
\centering \includegraphics[width=0.49\textwidth]{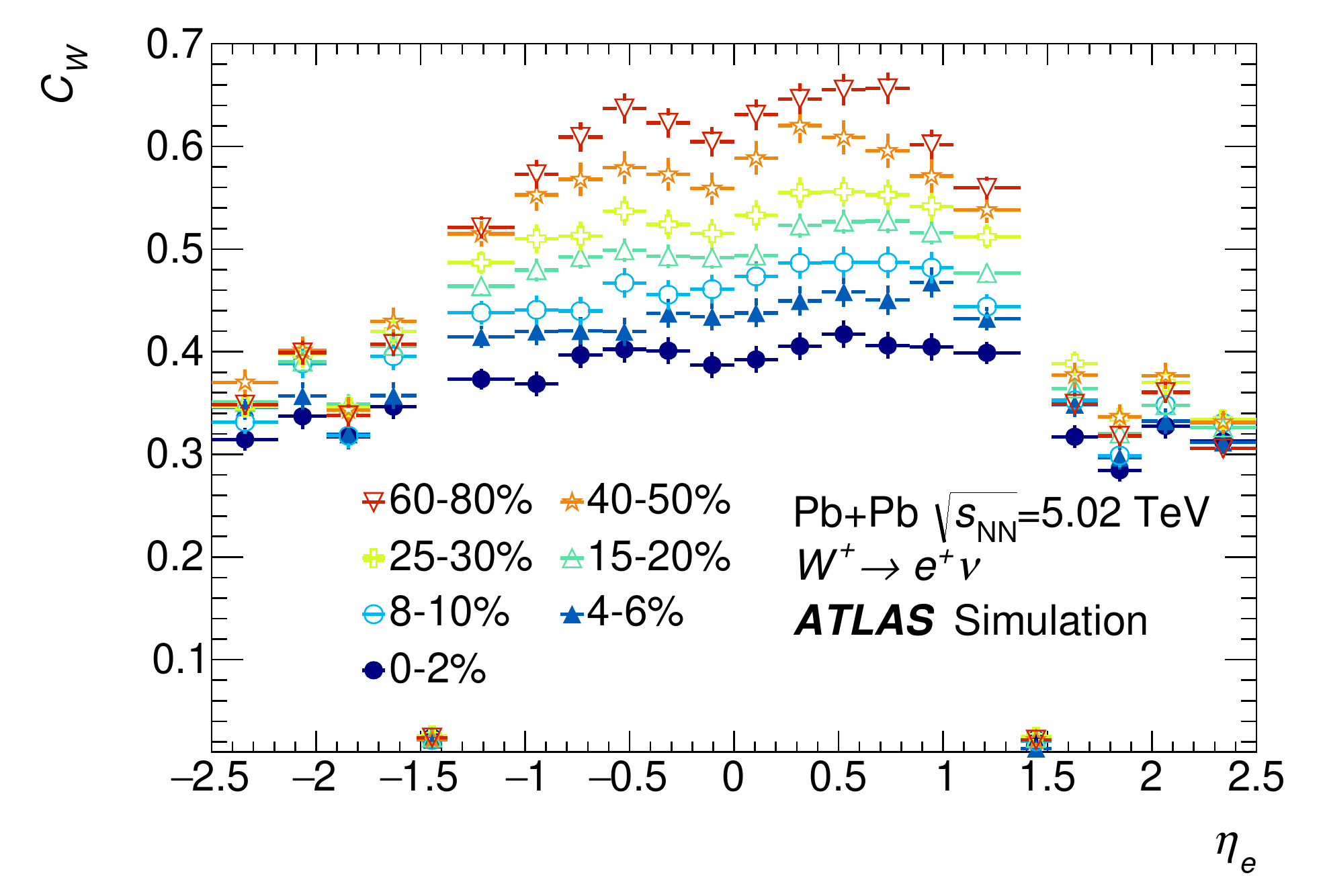}
\centering \includegraphics[width=0.49\textwidth]{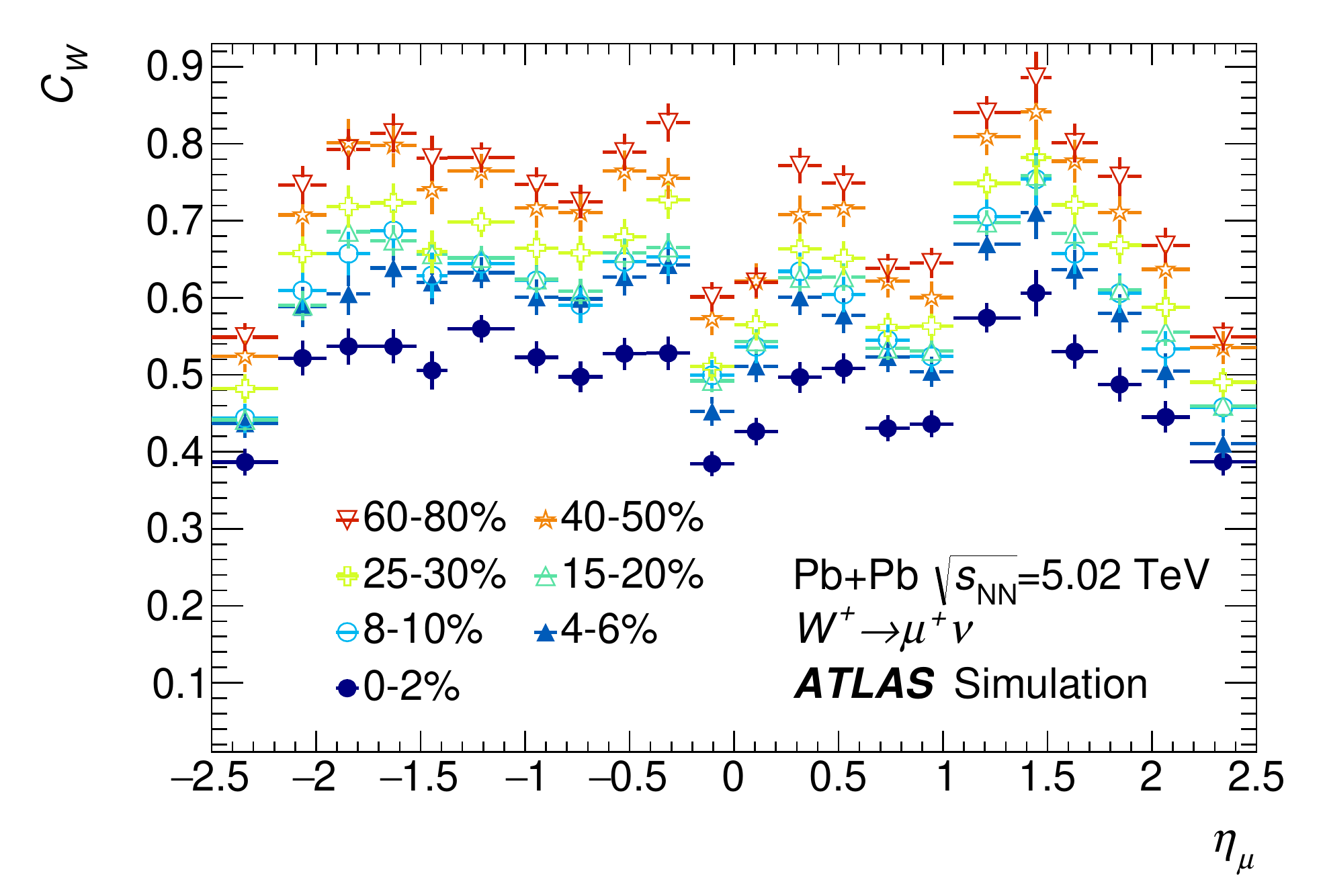}
\caption{ Correction factor \Cw\ for positive electrons (left) and positive muons (right) as a function of $\eta_\ell$ evaluated in selected centrality classes. The error bars represent the statistical uncertainties. }
\label{fig:cw_cent}
\end{figure}
 
Figure~\ref{fig:cw_cent} shows the \Cw\ correction factors evaluated for positive electrons and muons as a function of $\eta_\ell$ and centrality in events from selected centrality ranges. A sizeable evolution with event centrality is observed for both channels. The centrality dependence is mainly driven by the \ptmiss resolution which deteriorates with increasing event activity. The usage of the data overlay procedure in production of the MC samples ensures a good description of the underlying event in the simulation. The \ptmiss resolution is also tested with $Z\to\ell^+\ell^-$ events as a function of centrality, and reasonably good agreement between \ptmiss distributions in the data and MC simulation is found. Residual differences are due to a misalignment of the ID. They are covered by the systematic uncertainty discussed in Section~\ref{sec:systematics}. The veto on $Z\to\ell^+\ell^-$  decays also contributes to the change of the \Cw\ correction factor in the most central events, where the rate of `loose' quality leptons increases. The muon reconstruction and identification efficiencies are measured to be above 90\% and not dependent on centrality. The efficiency of the muon isolation selection is measured to be $\sim$90\% in the barrel region and $\sim$96\% in the endcap region. This selection was optimised as a function of centrality, and therefore no dependence on detector occupancy is observed. The electron isolation efficiency depends on centrality and varies from $\sim$90\% in peripheral events to  $\sim$75\% in the most central events. In the electron channel, a significant difference in the evolution of the \Cw\ correction factor can be noticed between the central~($|\eta|<1.37$) and forward pseudorapidities~($1.52<|\eta|<2.47$). That behaviour can be attributed to the electron reconstruction efficiency, which increases in the forward region as a function of centrality from $\sim$75\% to $\sim$95\% almost compensating for other effects, while in the barrel region it changes from $\sim$90\% to $\sim$95\%. The increase in the reconstruction efficiency is caused by the increasing number of charged-particle tracks and a loose requirement on matching the track to the EM cluster. Finally, the electron identification is optimised to have a constant efficiency as a function of centrality and its value is above 80\%. For negative electrons and muons, the \Cw\ correction factors (not shown in the figure) are sensitive to the same effects and result in similar behaviour as a function of $\eta_\ell$ and centrality.
 
As shown in Eq.~\eqref{eq:sigmaNN}, particle production yields in heavy-ion collisions are often presented in terms of the number of counts per MB collision. The total number of MB collisions corresponding to the analysed dataset is extracted from a MB data sample as described in Ref.~\cite{HION-2016-07} and is equal to $2.99 \times 10^9$ collisions for the 0--80\% centrality class. The \Nevt\ values for centrality classes used in this analysis are derived as corresponding fractions of this number.
 
\subsection{Systematic uncertainties}
\label{sec:systematics}
Systematic uncertainties of the measured observables are determined separately for electron and muon decay channels as well as for positive and negative lepton electric charges. They are estimated for each pseudorapidity bin and centrality class. The sources of systematic uncertainties considered are described below.
 
The uncertainties in the measurement of lepton reconstruction, identification, trigger and isolation efficiency scale factors are separated into statistical and systematic components. The statistical uncertainties of the scale factors are propagated to the final results using a toy MC approach, while the systematic uncertainties are propagated in a fully correlated way across all lepton $\eta_\ell$ and $\pt^\ell$ bins. In the electron channel, the largest of these uncertainties is related to the electron identification efficiency measurement, which is limited by statistical precision and is at most 4\% for yields measured as a function of $\eta_\ell$. The dominant uncertainty in the muon channel comes from the measurement of muon trigger efficiency and varies between 2\% and 4\% for yields measured as a function of $\eta_\ell$.
 
Systematic uncertainties related to electron energy calibration and muon momentum calibration are evaluated by varying scale and resolution corrections applied to the lepton candidates. The corresponding variations of measured yields are $\sim$0.1\% and their contribution to the systematic uncertainty is, therefore, neglected.

The resolution and background rejection power of the reconstructed \ptmiss\ depends on the contribution from low-\pt\ particles produced in the underlying event. In order to assess the impact of this contribution, the threshold for the \pt\ of ID tracks used in the \ptmiss\ calculation is varied in both the data and MC simulation from its nominal value of 4~\GeV\ up and down by 1~\GeV, and the full analysis is repeated.  The higher track \pt\ threshold  minimises the impact of the underlying event on the \ptmiss\ resolution but also removes tracks required to balance the transverse energy of the event. Studies performed in MB events showed that the 3~\GeV\ \pt\ threshold introduces a bias in the mean values of the $x$ and $y$ components of \ptmiss. Therefore, the lower track \pt\ threshold introduces sources of spurious \ptmiss\ related to a larger contribution of soft particles from the underlying event.
In addition, misalignment of the ID produces a charge-dependent bias in the measured \pt\ of tracks, which is specific to the analysed dataset. The bias is evaluated to be 2\% for tracks with a \pt\ of about 40~\GeV. Since the signal lepton track drives the value of reconstructed \ptmiss, an uncertainty due to this bias is evaluated by varying the \ptmiss\ scale in data by $\pm$ 2\%. This variation is applied in a correlated way to events with leptons of positive and negative charge. If an anti-correlated variation is used instead, the impact on the charge asymmetry measurement is found to be negligible. The total uncertainty in measured yields due to the \ptmiss\ reconstruction and ID misalignment varies as a function of centrality between 2\% and 4\% for the electron channel and between 1\% and 3\% for the muon channel.

An uncertainty varying from 2\% to 8\% in the electron channel and $\sim$2\% in the muon channel is associated with the data-driven estimation of the multi-jet background. The variation in the electron channel is $\eta$-dependent. A smaller effect is observed in the barrel region, while the 8\% variation is observed in the endcap region where the fraction of the multi-jet background is significant. This uncertainty is partially estimated using systematic variations of the correction applied to the shape of the multi-jet background template. By default, the distance, $d$, used in the shape reweighting procedure described in Section~\ref{sec:bkgr} is defined relative to the mean value of the signal isolation. However, the width of the signal isolation region in $\pt^\mathrm{iso}/\pt^\mu$ is non-negligible, and for systematic variations, $d$ is recalculated relative to the boundaries of this region. In addition, the shape of the multi-jet background template is found to be dependent on $\eta$. The uncertainty due to this effect is estimated by comparing templates constructed using leptons from separate $\eta$ regions (either barrel or endcap) to the nominal template. The contribution from residual differences in shapes to the total systematic uncertainty related to the multi-jet background estimation is 1--3\% in the electron channel and $\sim$1\% in the muon channel.
An additional uncertainty is related to the statistical precision of multi-jet background templates extracted from data. It is evaluated to be $\sim$2\% (barrel) and $\sim$4\% (endcap) in the electron channel and $\sim$2\% in the muon channel. The two independent components, representing systematic  and statistical contributions to the uncertainty, are added in quadrature while reporting the total uncertainty.
 
Uncertainties in the estimation of EW and top-quark backgrounds are evaluated by varying their normalisation within the error of their cross-sections. These variations result in up to 0.5\% uncertainty in the electron channel and up to 0.2\% uncertainty in the muon channel.
 
The average nuclear thickness function, \avTaa, is used to normalise the $W^\pm$ boson production yields. The \avTaa\ uncertainties listed in Table~\ref{tab:centrality} are used to vary the normalised yields. The resulting normalisation uncertainty ranges from 0.9\% in the most central collisions to 8.2\% in the 60--80\% centrality class, while for the 0--80\% centrality class, it is 1.6\%.
 
The simulated event samples produced for different isospin combinations of colliding nucleons are normalised assuming a fixed collision rate for each combination. The impact of this assumption is checked by correcting the data after background subtraction using \Cw\ correction factors evaluated from the signal simulation produced for either \pp\ collisions or $nn$ collisions. No significant difference is observed compared to the application of nominal corrections.
 
\begin{figure}[thb!]
\centering
\includegraphics[width=0.49\textwidth]{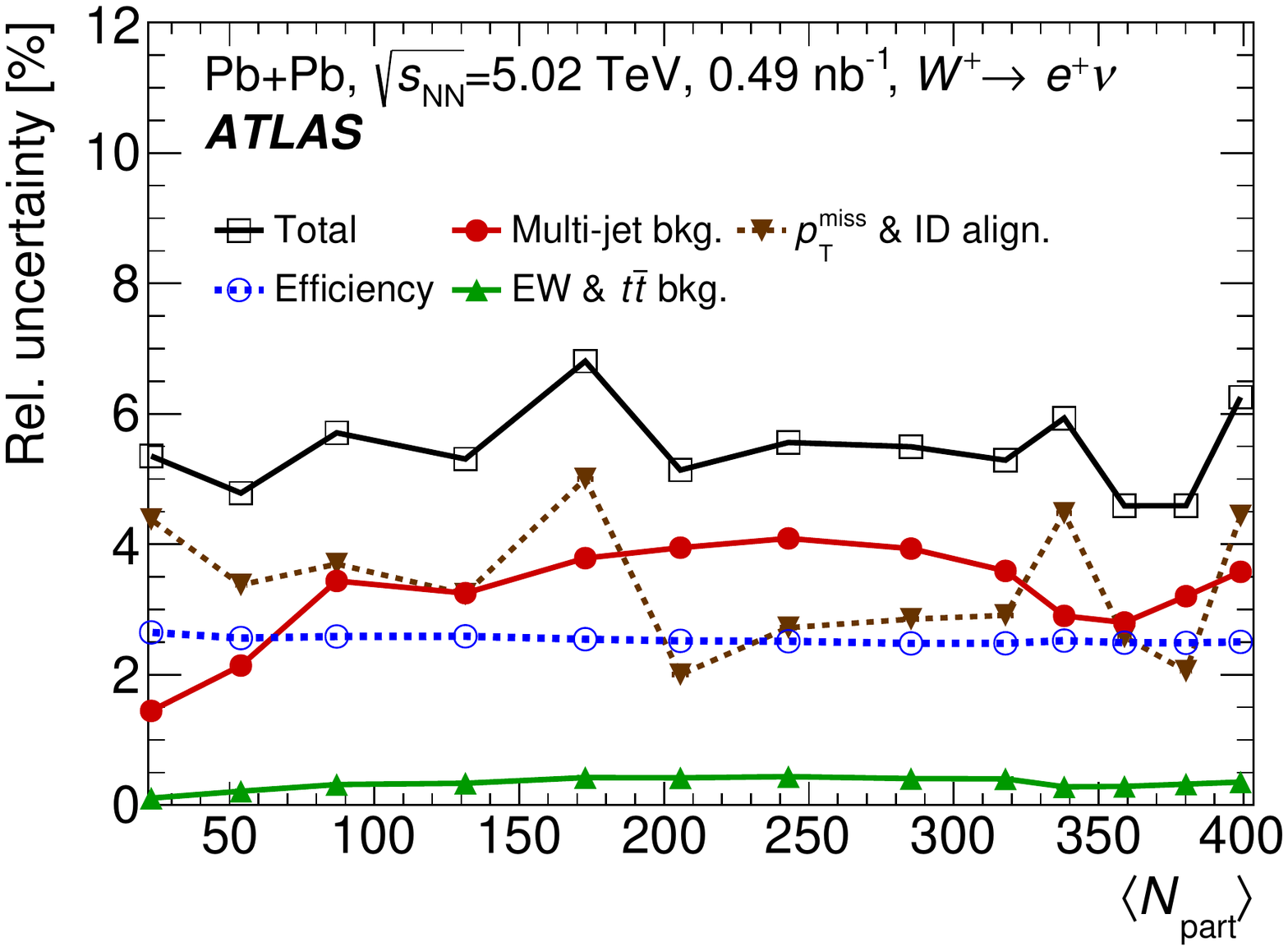}
\hfill
\includegraphics[width=0.49\textwidth]{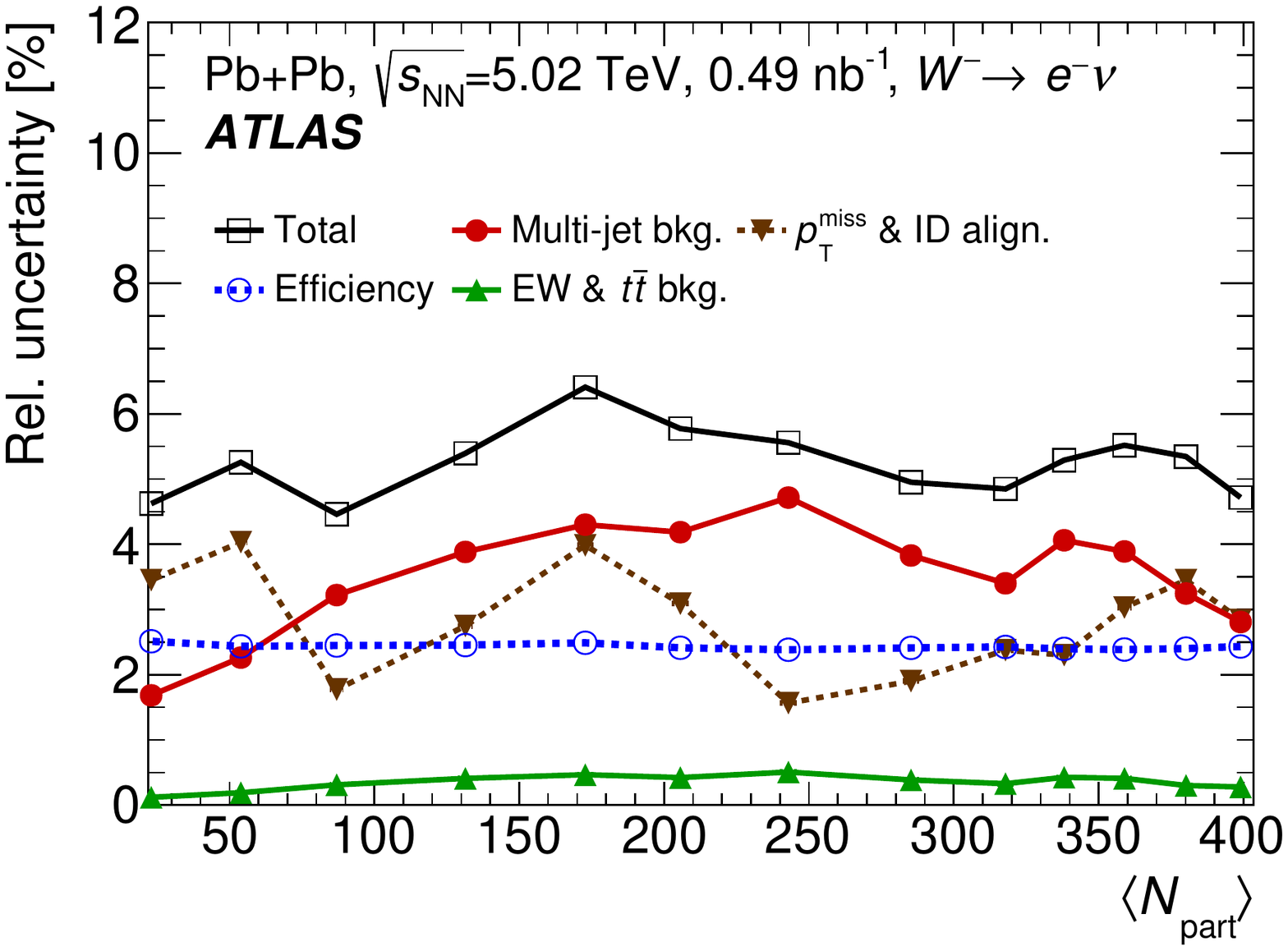}
\caption{Relative systematic uncertainties of $W^+$~(left) and $W^-$~(right) boson production yields measured in the electron decay channel evaluated as a function of \avNpart. The total systematic uncertainty is represented by open squares, while other markers represent contributions from individual sources of uncertainty. The uncertainties related to electron efficiency corrections~(``Efficiency''), \ptmiss\ reconstruction and ID misalignment uncertainties~(``\ptmiss\ \& ID align.''), as well as the uncertainties related to the estimation of EW and top-quark backgrounds~(``EW \& $t\bar{t}$ bkg.''), are added in quadrature.}
\label{fig:sys_total_Npart_ele}
\end{figure}

\begin{figure}[thb!]
\centering
\includegraphics[width=0.49\textwidth]{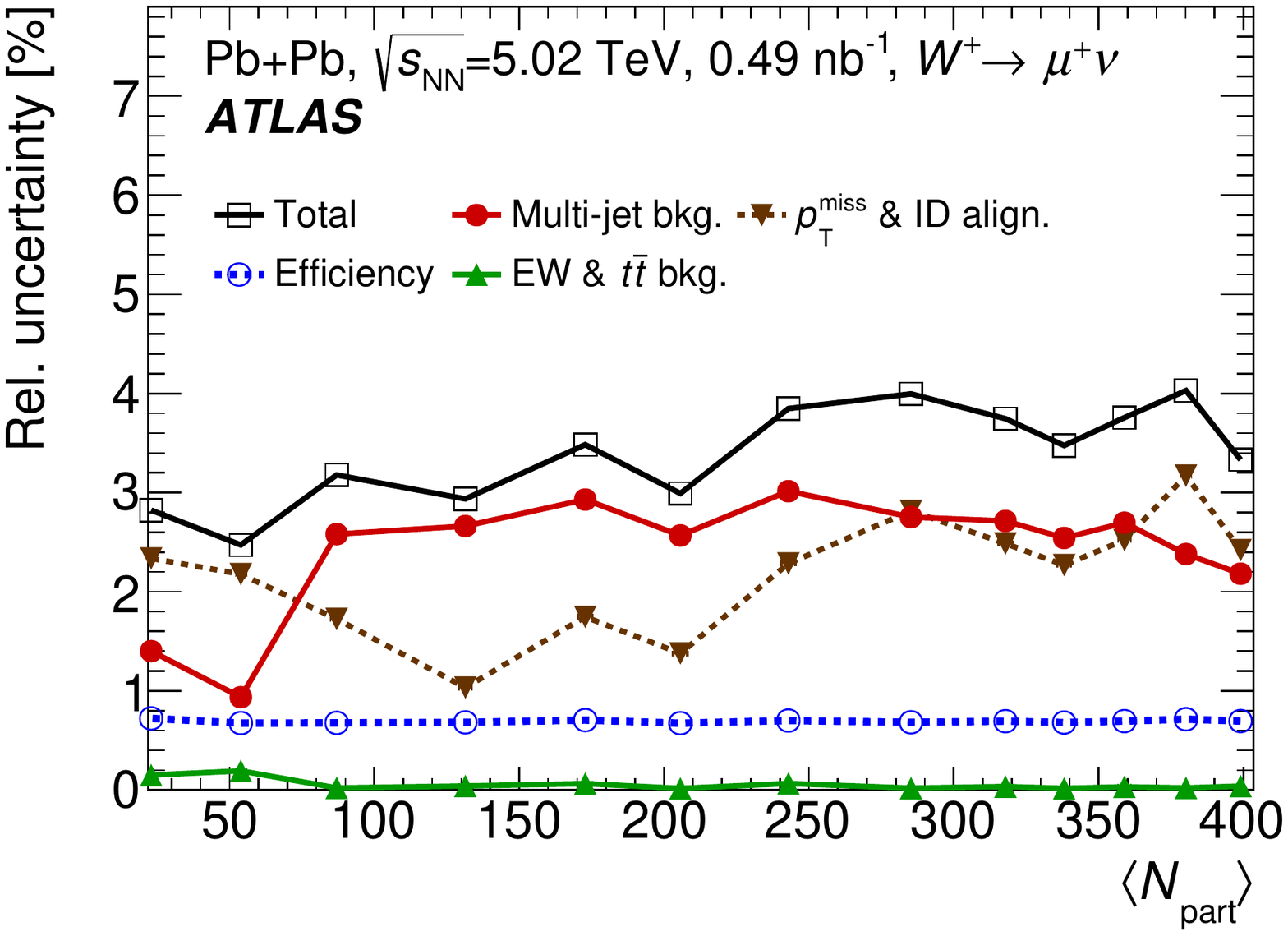}
\hfill
\includegraphics[width=0.49\textwidth]{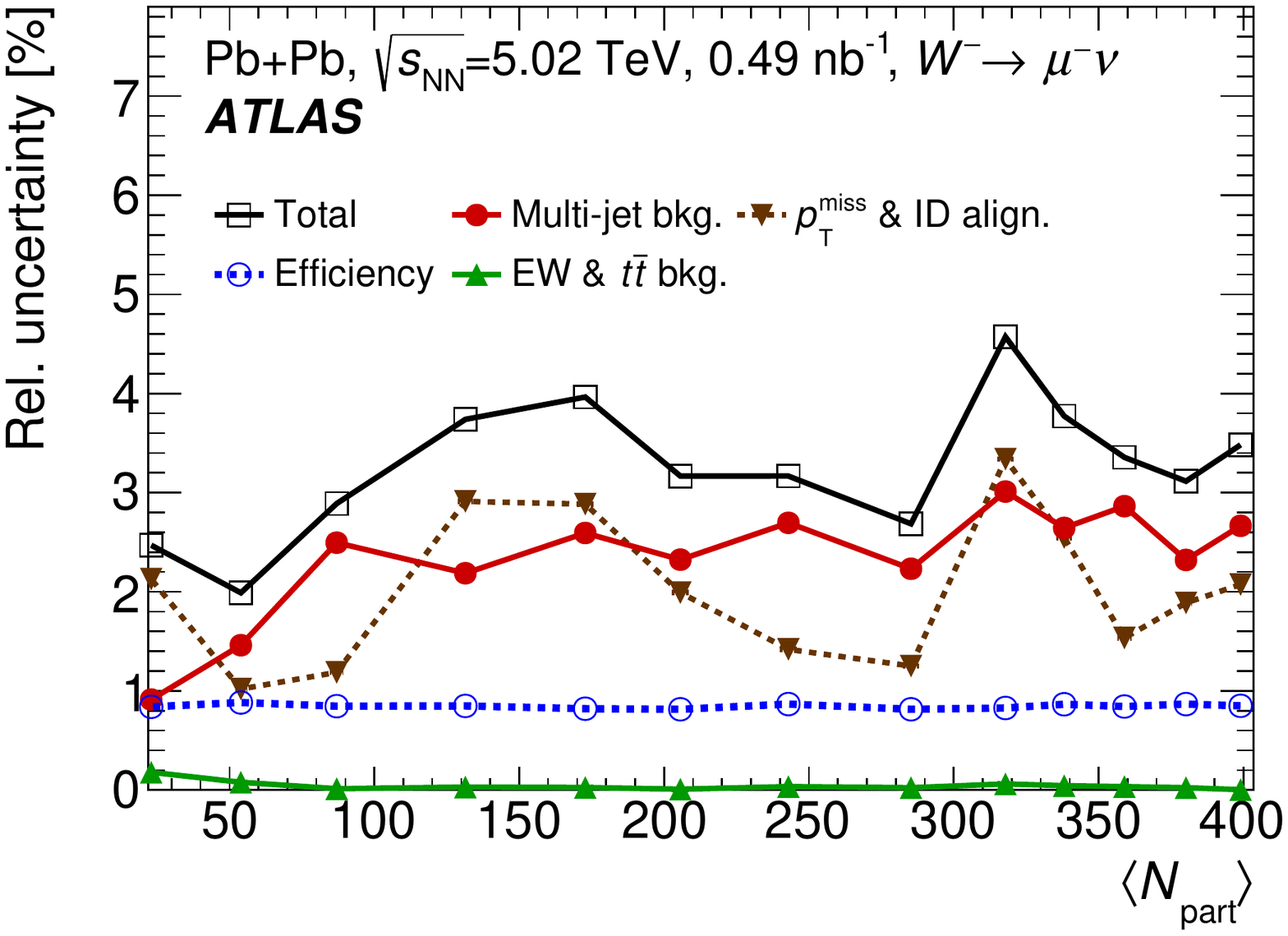}
\caption{Relative systematic uncertainties of $W^+$~(left) and $W^-$~(right) boson production yields measured in the muon decay channel evaluated as a function of \avNpart. The total systematic uncertainty is represented by open squares, while other markers represent contributions from individual sources of uncertainty. The uncertainties related to muon efficiency corrections~(``Efficiency''), \ptmiss\ reconstruction and ID misalignment uncertainties~(``\ptmiss\ \& ID align.''), as well as the uncertainties related to the estimation of EW and top-quark backgrounds~(``EW \& $t\bar{t}$ bkg.''), are added in quadrature.  }
\label{fig:sys_total_Npart_mu}
\end{figure}

A summary of systematic uncertainties as a function of \avNpart\ is shown in Figures~\ref{fig:sys_total_Npart_ele} and~\ref{fig:sys_total_Npart_mu} for the electron and muon channels, respectively.
The total systematic uncertainty of the $W^\pm$ boson yields obtained in the electron decay channel varies as a function of $\eta_\ell$ in the barrel region between 4\% and 5\%. This uncertainty is affected by the statistical precision of the efficiency scale factors measured in bins of $\eta_\ell$. Total systematic uncertainties in the endcap regions are much larger and reach 10\%. This is caused by the significant multi-jet background contamination and significantly lower statistical precision of the efficiency scale factor estimation. The total systematic uncertainty for $\eta_\ell$-integrated yields is largely independent of centrality and amounts to about 5\%. In the muon channel, the precision of the measurement for all $\eta_\ell$ values is similar to the precision of the electron channel measurement in the barrel region. As a function of centrality, the total systematic uncertainty in the muon channel is approximately constant at about 3\%.
 
\FloatBarrier
\subsection{Channel combination}
\label{sec:combination}
The corrected electron and muon channel measurements are combined using the Best Linear Unbiased Estimate (BLUE) method~\cite{Valassi:2003mu}, accounting for correlations of systematic uncertainties across the channels and measurement bins.
For some systematic errors, no covariance matrix is available, and therefore some assumptions about correlations between bins and production channels need to be made. The \avTaa\ uncertainty and theoretical uncertainty in the background production cross-sections for simulated processes are assumed to be fully correlated between bins and channels. This approach is justified as they are used as simple normalisation factors which are common to both channels and across all analysis bins. The uncertainties in the multi-jet background estimation are assumed to be fully correlated between analysis bins and uncorrelated between the decay channels. Finally, the uncertainties related to the \ptmiss\ reconstruction are conservatively assumed to be correlated between both the analysis bins and the decay channels.
% End of text imported from the .//Sec_4.tex input file
% The next lines are included from the .//Sec_5.tex input file
\section{Results}
\label{sec:results}
\begin{figure}[!tbh]
\centering
\includegraphics[width=0.49\textwidth]{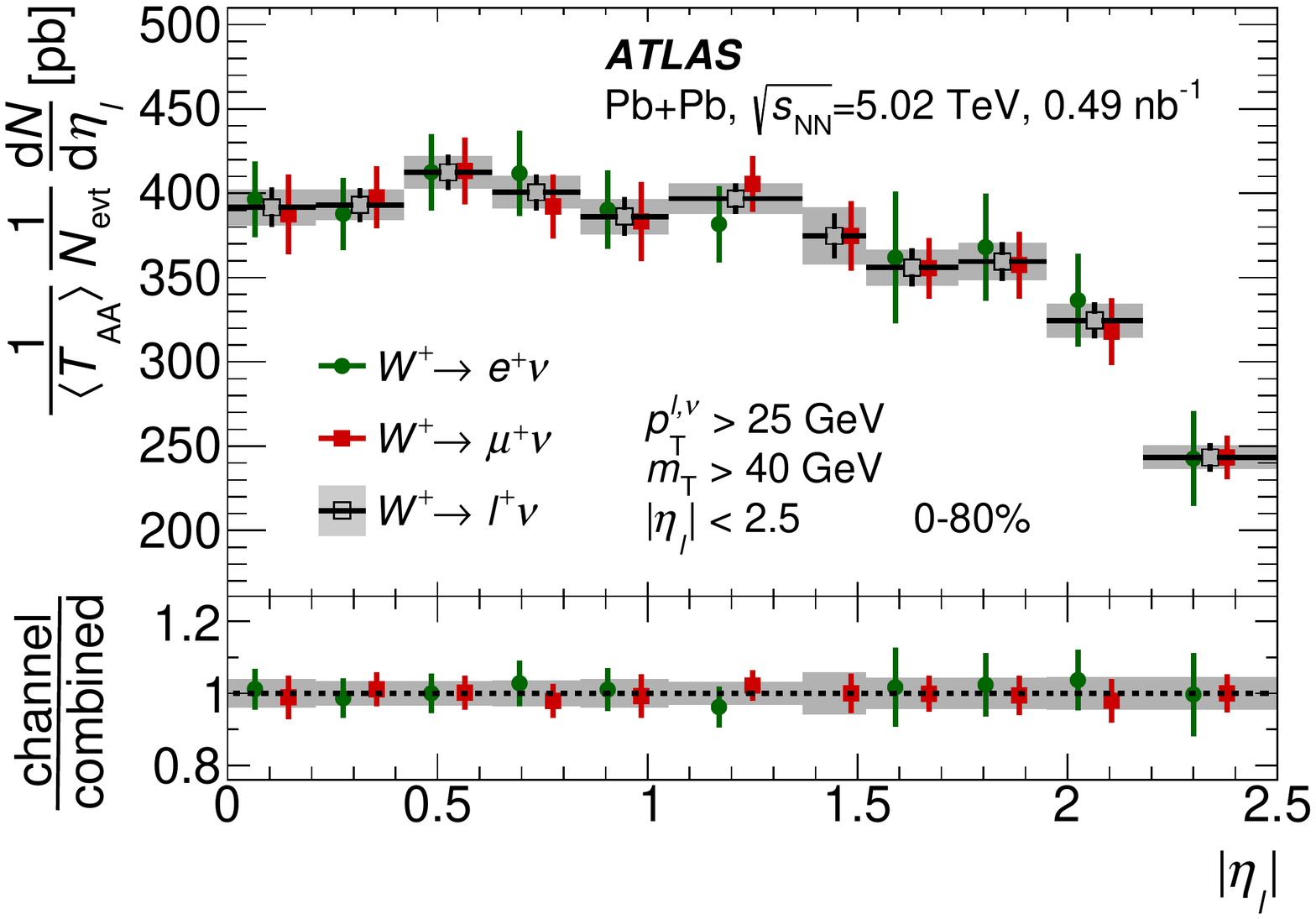}
\hfill
\includegraphics[width=0.49\textwidth]{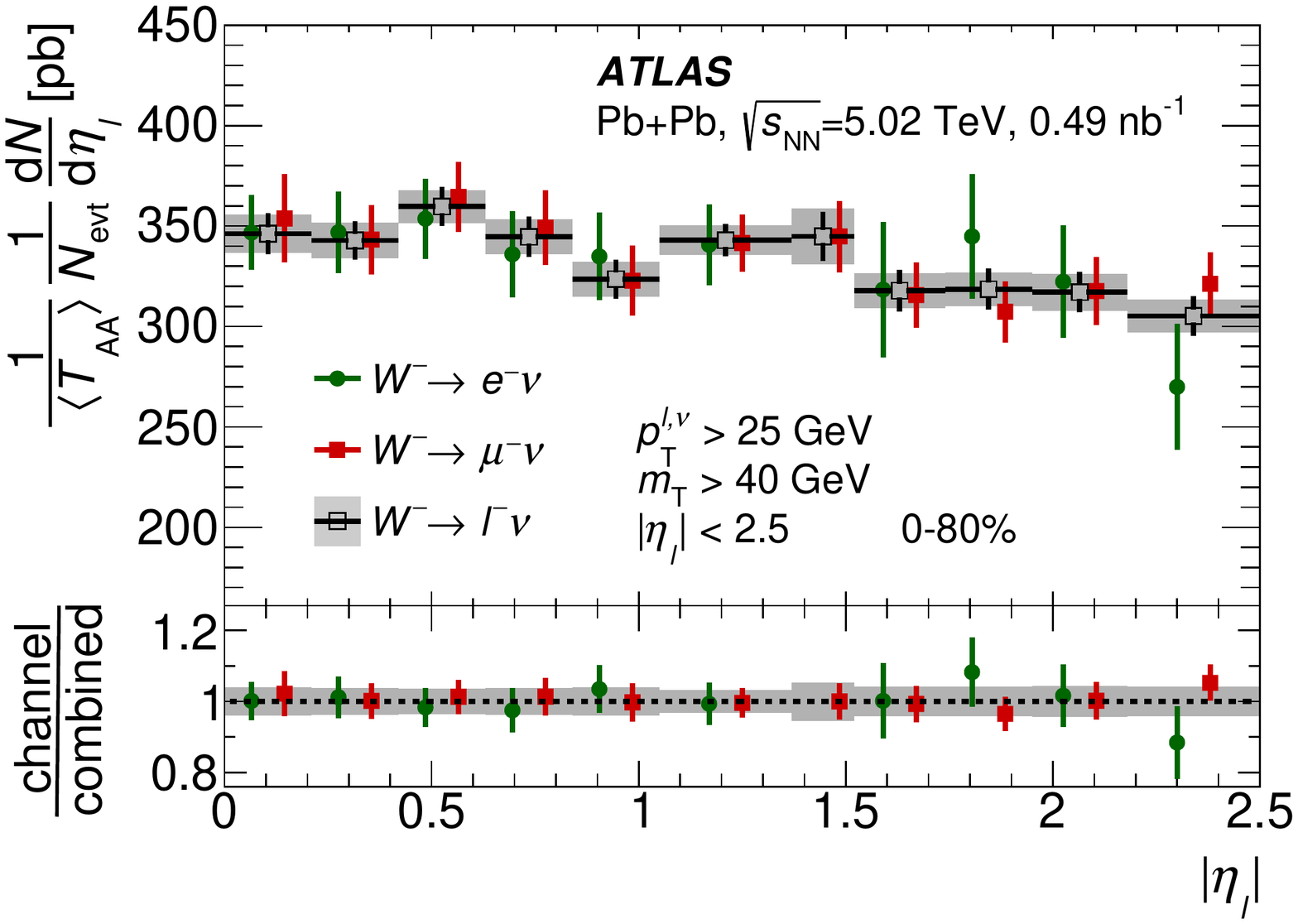}
\caption{Differential normalised production yields for $W^+$~(left) and $W^-$~(right) bosons as a function of absolute pseudorapidity of the charged lepton shown separately for electron and muon decay channels as well as for their combination. Statistical and systematic uncertainties of the combined yields are shown as bars and shaded boxes, respectively. For the individual channels, only the total uncertainties are shown as error bars. Systematic uncertainties related to \avTaa\ are not included. The lower panels show the ratios of channels to combined yields in each bin with error bars and shaded boxes representing the total uncertainties of the channels and combined yields, respectively. The points for individual channels are shifted horizontally for better visibility.}
\label{fig:eta-comparison}
\end{figure}

Figure~\ref{fig:eta-comparison} shows a comparison between the differential normalised production yields for  $W^+$ and $W^-$ bosons obtained for the electron and muon decay channels as a function of the absolute value of the charged-lepton pseudorapidity, $|\eta_\ell|$. The combined dataset is also shown on the same figure. Good agreement is found between the two decay modes, supporting the combination of the measurements. The distribution for $W^+$ bosons falls steeply at large $|\eta_\ell|$, whereas for $W^-$ bosons, it tends to be flat with $|\eta_\ell|$. This is attributed to the fact that high-\pt $W^\pm$ bosons are mostly left-handed~\cite{Berger:2009ep} and preferentially produced in the valence-quark direction, thus towards non-zero pseudorapidity. The $W^+$ boson decays into a right-handed positive lepton, which is thus boosted back towards lower $|\eta_\ell|$, while the $W^-$ boson decays into a left-handed negative lepton which is boosted towards higher $|\eta_\ell|$.

\begin{figure}[!tbh]
\centering
\includegraphics[width=0.49\textwidth]{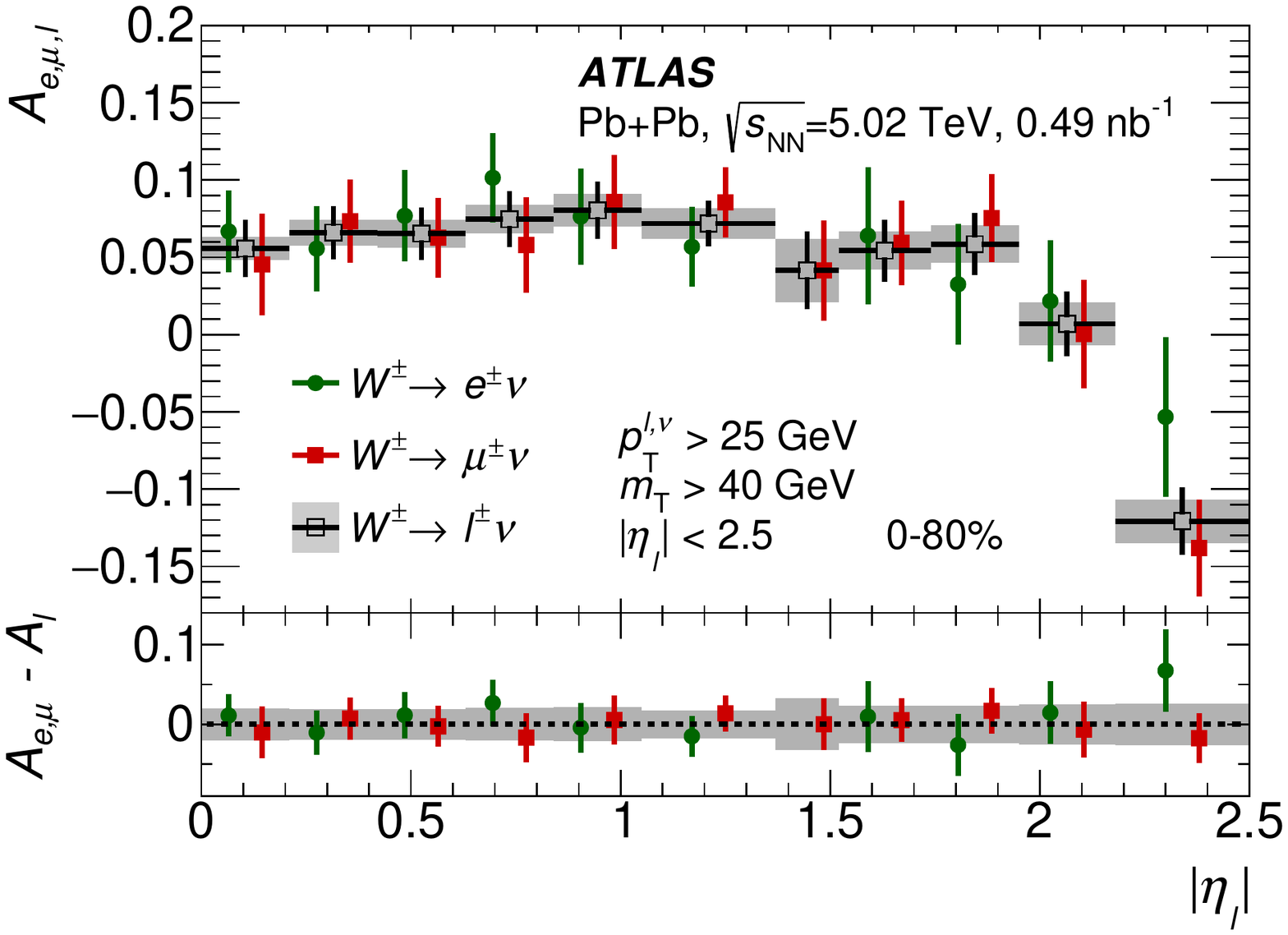}
\caption{Lepton charge asymmetry as a function of absolute pseudorapidity of the charged lepton, $|\eta_\ell|$, measured for $W^\pm$ bosons decaying into electrons and muons as well as for the combination of the two channels. Statistical and systematic uncertainties of the combined charge asymmetry are shown as bars and shaded boxes, respectively. For the individual channels, only the total uncertainties are shown as error bars. The lower panel shows the differences between the asymmetries measured for each channel separately and their combination with error bars and shaded boxes representing the total uncertainties of the channels and combined asymmetry, respectively. The points for individual channels are shifted horizontally for better visibility.}
\label{fig:charge-comparison}
\end{figure}

Figure~\ref{fig:charge-comparison} shows a comparison between lepton charge asymmetries obtained for the electron and muon decay channels as a function of the charged-lepton absolute pseudorapidity. Good agreement is found between the two decay modes, which supports the combination of the two datasets. The resulting combined data points are also shown on the same figure.
 
\begin{figure}[!tbh]
\centering
\includegraphics[width=0.49\textwidth]{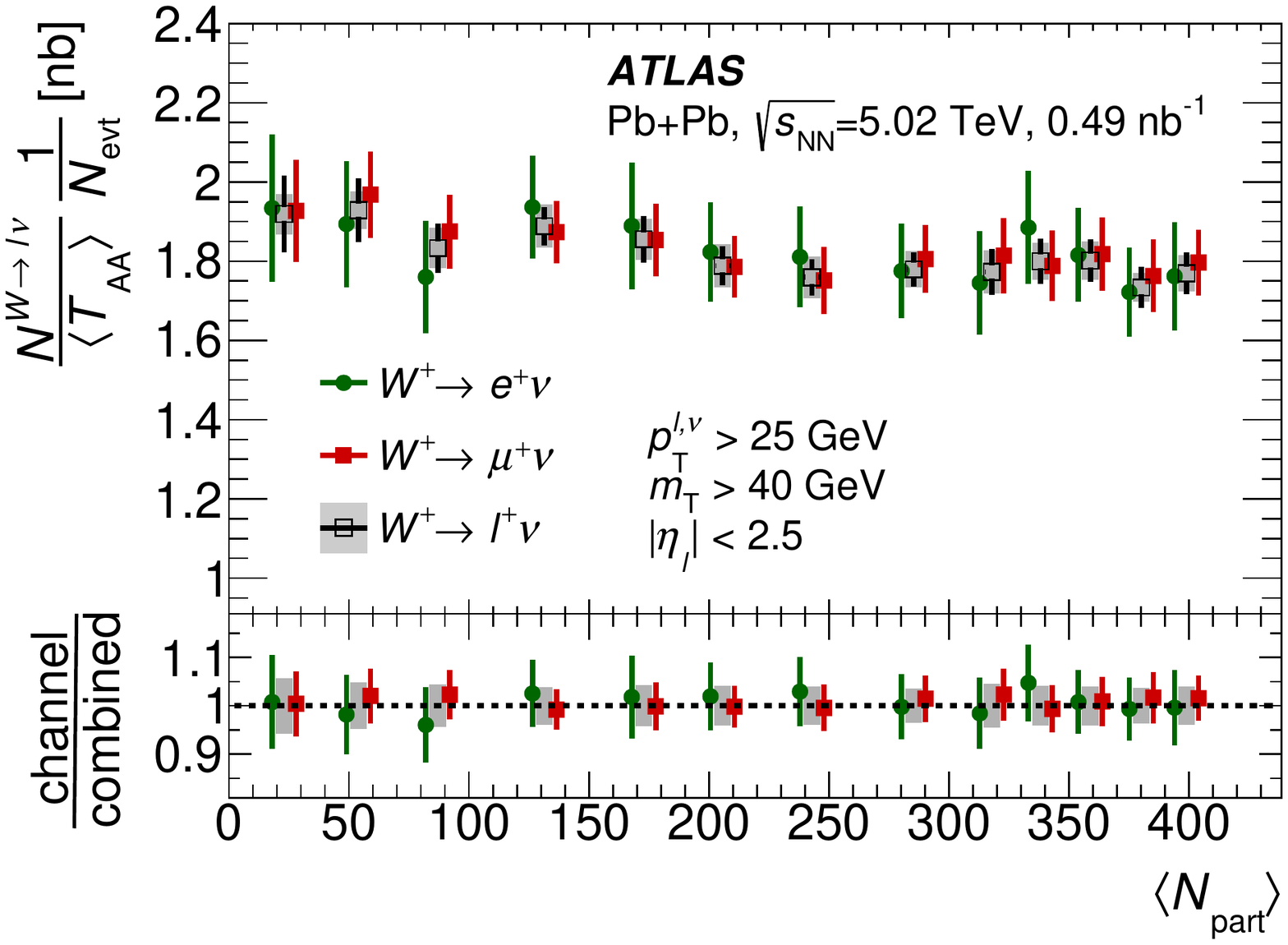}
\hfill
\includegraphics[width=0.49\textwidth]{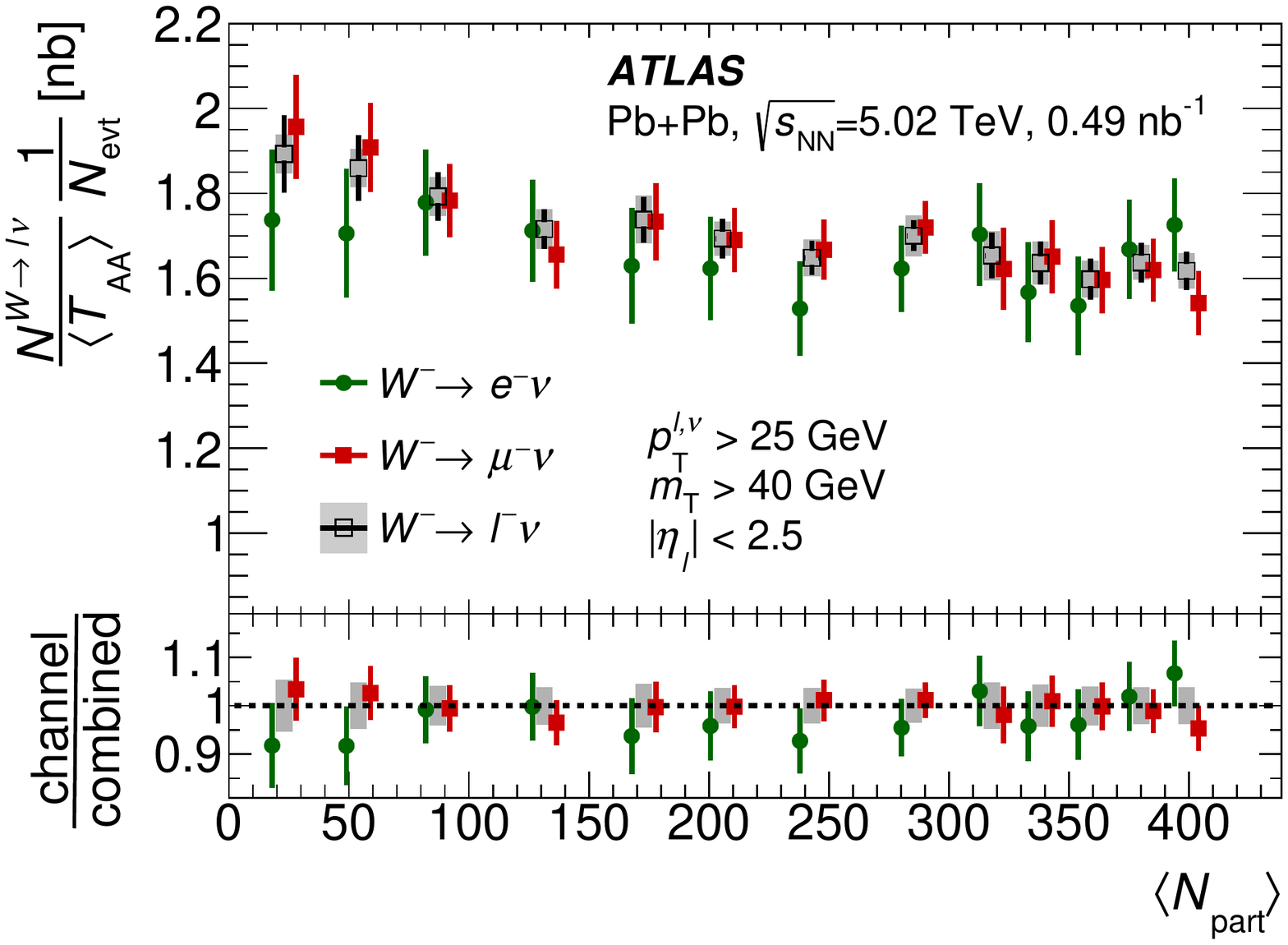}
\caption{Normalised production yields for $W^+$~(left) and $W^-$~(right) bosons as a function of \avgNpart\ shown separately for electron and muon decay channels as well as for their combination. Statistical and systematic uncertainties of the combined yields are shown as bars and shaded boxes, respectively. For the individual channels, only the total uncertainties are shown as error bars. Systematic uncertainties related to \avTaa\ are not included. The lower panels show the ratios of channels to combined yields in each bin with error bars and shaded boxes representing the total uncertainties of the channels and combined yields, respectively. The points for individual channels are shifted horizontally for better visibility.}
\label{fig:centrality-comparison}
\end{figure}

Figure~\ref{fig:centrality-comparison} shows a comparison of the normalised production yields for $W^+$ and $W^-$ bosons obtained for the electron and muon decay channels, as well as their combination, as a function of the event centrality (represented by  \avgNpart). Here, also, good agreement between the two decay modes is observed.
 
\begin{figure}[!tbh]
\centering
\includegraphics[width=0.49\textwidth]{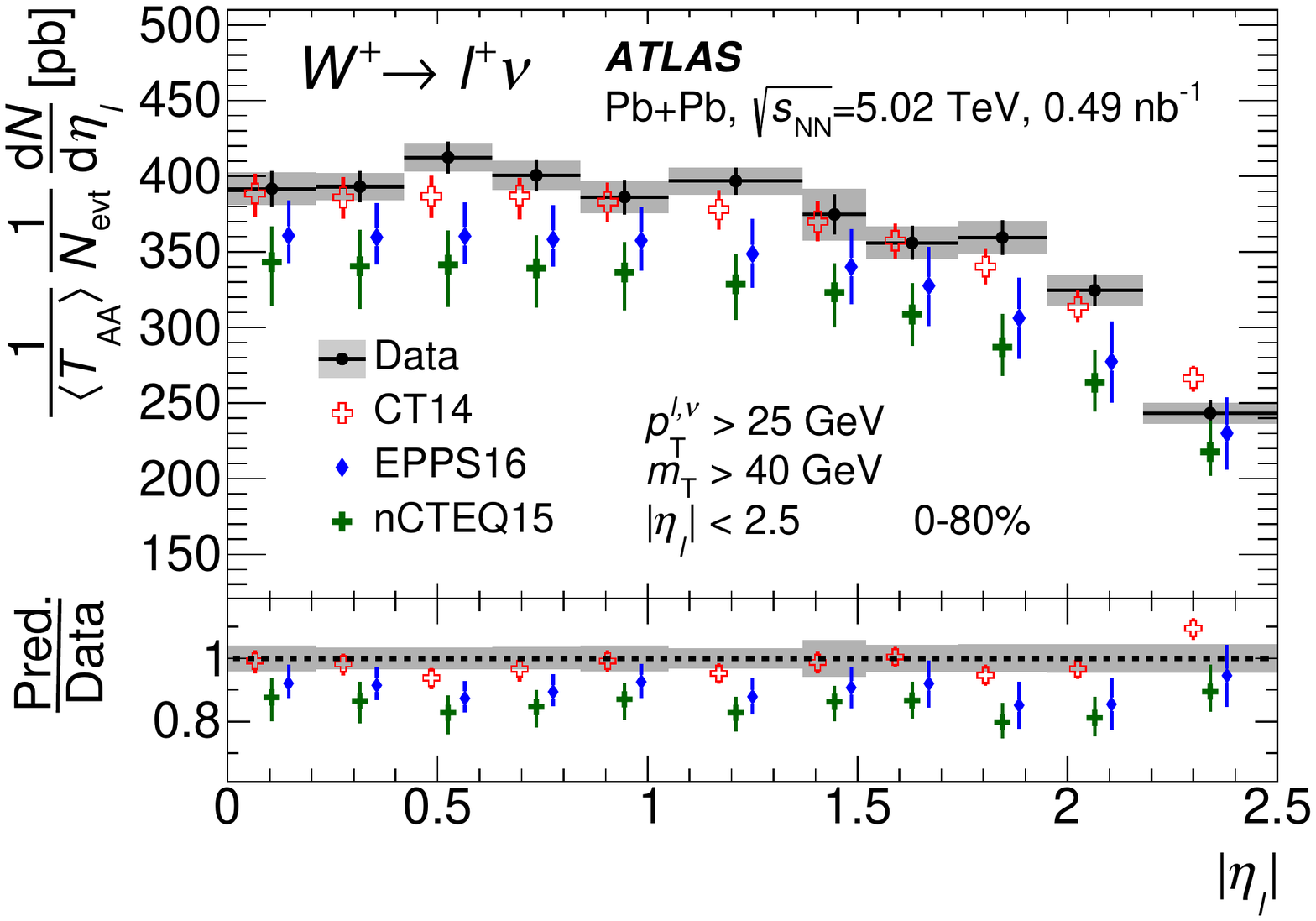}
\hfill
\includegraphics[width=0.49\textwidth]{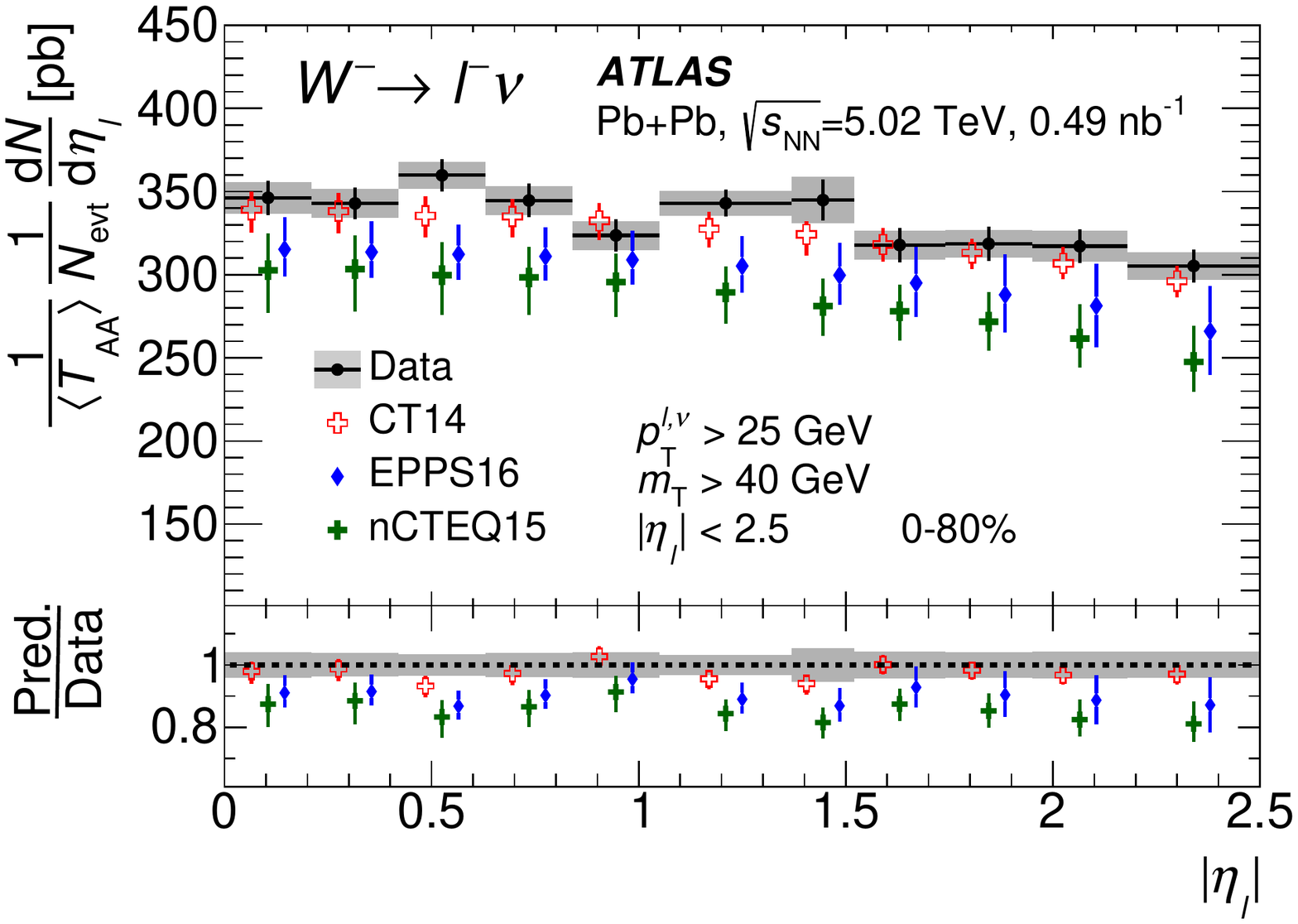}
\caption{Differential normalised production yields for $W^+$~(left) and $W^-$~(right) bosons as a function of absolute pseudorapidity of the charged lepton for the combined electron and muon channels. Error bars show statistical uncertainties, whereas systematic uncertainties are shown as shaded boxes. Systematic uncertainties related to \avTaa\ are not included. The measured distributions are compared with theory predictions calculated with the \textsc{CT14 NLO} PDF set as well as with EPPS16 and nCTEQ15 nPDF sets. For the theory predictions, the error bars represent total uncertainties due to PDF uncertainties, scale variations and \alphas\ variations. The lower panels show the ratios of predicted yields to the measured ones, and the shaded band shows the sum in quadrature of statistical and systematic uncertainties of the data. The points for theory predictions are shifted horizontally for better visibility.}
\label{fig:result-eta}
\end{figure}

Figure~\ref{fig:result-eta} shows a comparison of combined differential normalised production yields for $W^+$ and $W^-$ bosons with theoretical predictions as a function of charged-lepton pseudorapidity. The predictions are calculated using the MCFM code~\cite{Campbell:2015qma} at NLO accuracy in QCD. The calculations are performed using either the free-nucleon \textsc{CT14 NLO} PDF set or one of two PDF sets including nuclear modifications~(nPDFs): EPPS16 or nCTEQ15. All predictions account for the isospin effect. Uncertainties in the theoretical predictions include contributions from PDF uncertainties, variations of the renormalisation and factorisation scales and variations of the strong coupling constant \alphas.
All predictions provide a good description of the shapes of the measured $|\eta_\ell|$ distributions.  The prediction based on the \textsc{CT14 NLO} PDF set differs by 2--3\% in normalisation compared with the data, while the predictions based on nPDFs underestimate the measured yields by 10--20\%. It should also be noted that the $W^+$~($W^-$) boson production cross-sections measured in the \pp\ system~\cite{HION-2018-02-witherratum} are larger by 5\% (4\%) than the CT14 NLO theory predictions for \pp\ collisions, corresponding to a difference of about one standard deviation.

\begin{figure}[!tbh]
\centering
\includegraphics[width=0.49\textwidth]{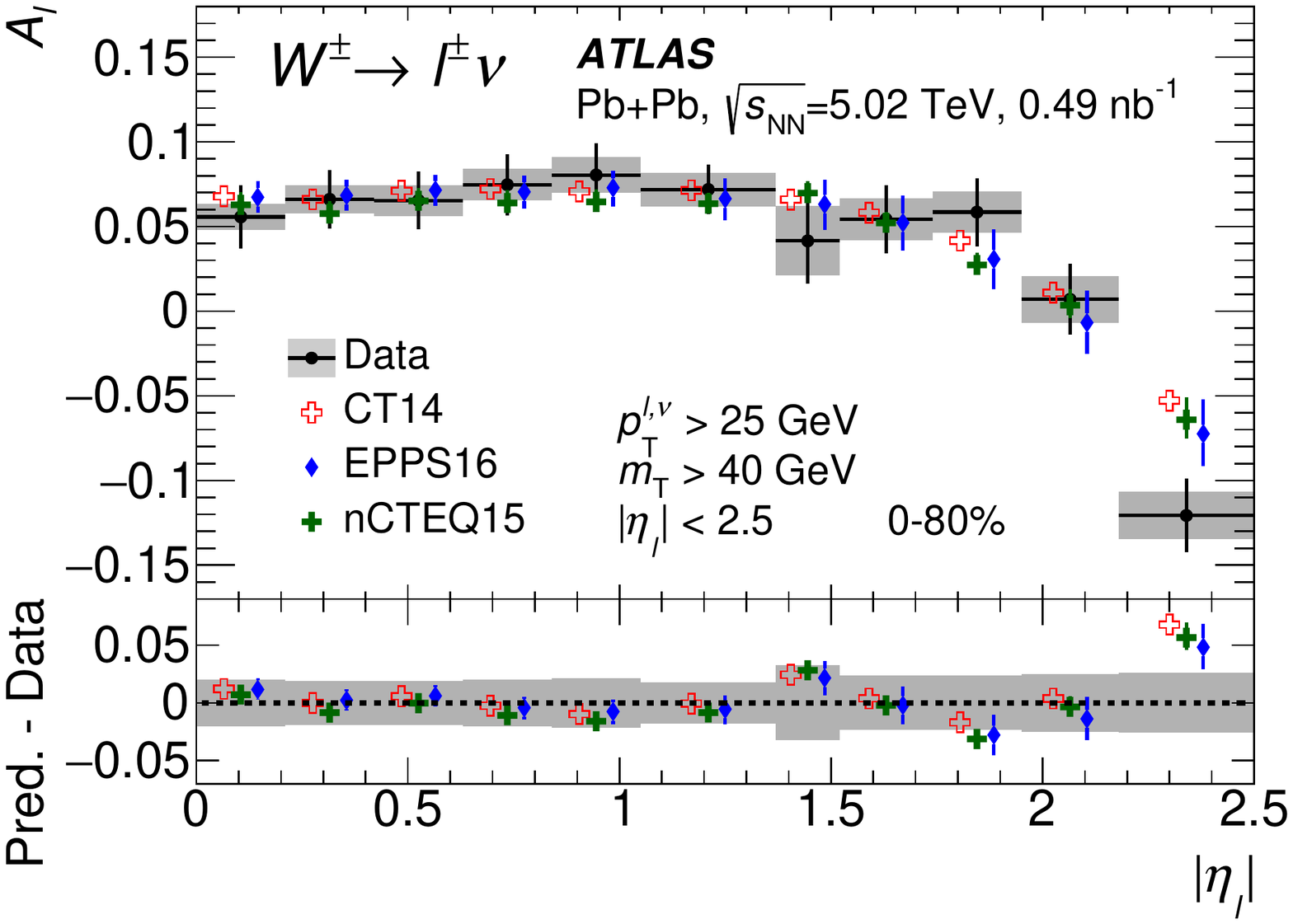}
\caption{Combined result for lepton charge asymmetry compared with theory predictions calculated with the \textsc{CT14 NLO} PDF set as well as with EPPS16 and nCTEQ15 nPDF sets. Error bars on the data points show statistical uncertainties, whereas systematic uncertainties are shown as shaded boxes. For the theory predictions, the error bars represent total uncertainties due to PDF uncertainties, scale variations and \alphas\ variations. The lower panel shows the differences between the predicted asymmetries and the measured ones with the shaded boxes representing the total experimental uncertainties. The points for theory predictions are shifted horizontally for better visibility.}
\label{fig:result-charge}
\end{figure}

The combined lepton charge asymmetry is compared with theoretical predictions in Figure~\ref{fig:result-charge}. All three predictions agree with the data within systematic uncertainties, except for the most forward $|\eta_\ell|$ bin. The isospin effect, which yields a larger fraction of $W^-\to\ell^-\nu$ events in \PbPb\ compared to \pp\ collisions in the forward region, results in a sign-change of the asymmetry that is observed within the $|\eta_\ell|$ acceptance of the measurement.
 
\begin{figure}[tbh!]
\centering
\includegraphics[width=0.49\textwidth]{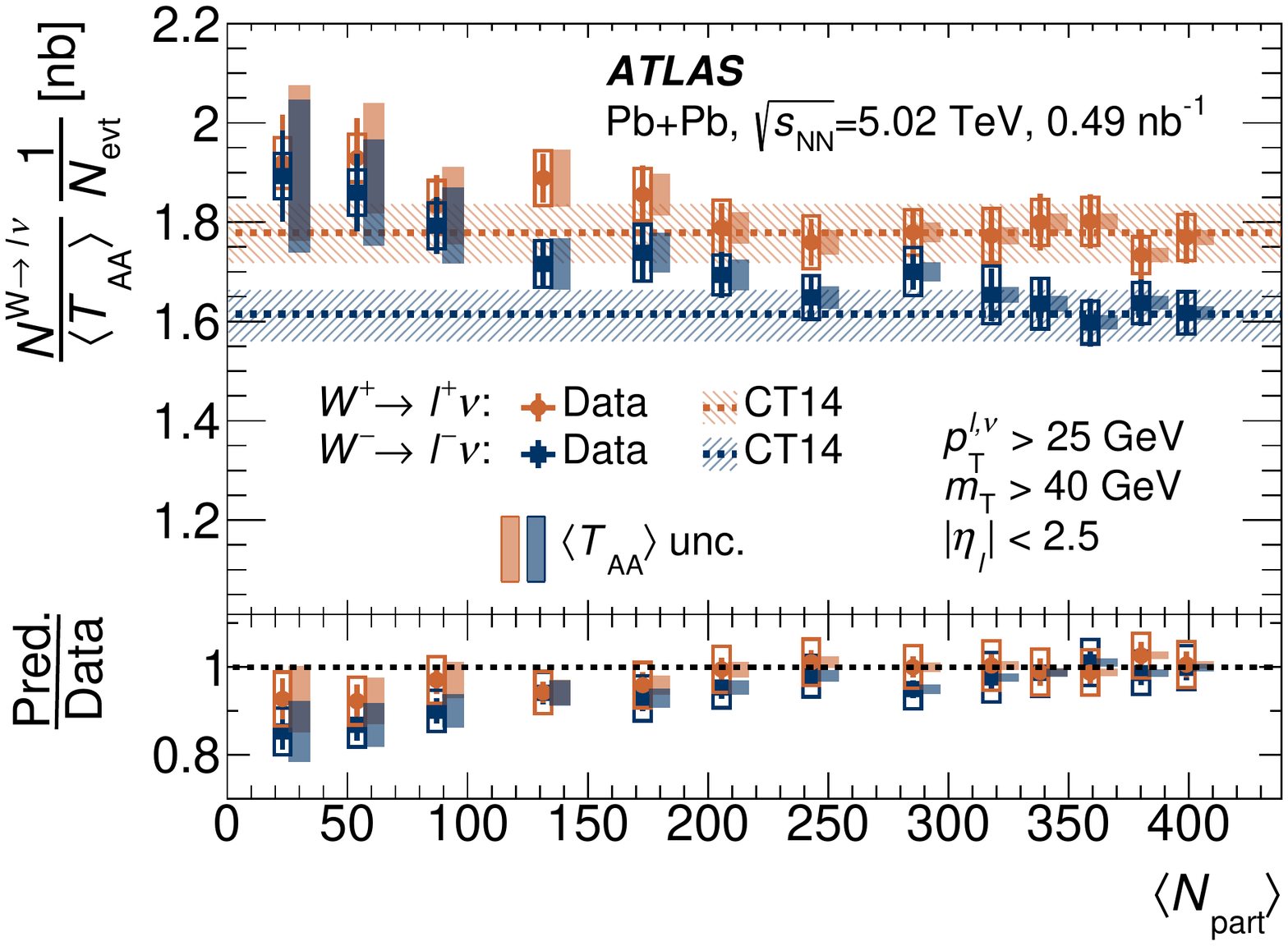}
\caption{Normalised production yields of $W^+$ and $W^-$ bosons as a function of \avgNpart\ shown for the combination of electron and muon decay channels. Predictions calculated using the \textsc{CT14 NLO} PDF set are shown as the horizontal bands. Error bars show statistical uncertainties, whereas systematic uncertainties are shown as the boxes around the data points. The systematic uncertainties due to \avTaa\ are not included in those boxes, and are shown as separate shaded boxes plotted to the right of the data points for better visibility. In the lower panel the ratios of the predictions to the measured yields are displayed, and the boxes around the data points show the sum in quadrature of statistical and systematic uncertainties of the data.}
\label{fig:result-centrality}
\end{figure}

Figure~\ref{fig:result-centrality} compares the normalised production yields of $W^+$ and $W^-$ bosons as a function of \avgNpart\ for the combined electron and muon channels. The normalised production yields for $W^+$ bosons are about 10\% higher than the yields for $W^-$ bosons. The data are also compared with theoretical predictions based on the CT14 NLO PDF set, which include the isospin effect.
The normalised production yields for $W^\pm$ bosons do not change with \avgNpart\ for mid-central and central collisions represented by \avgNpart\ values above 200. In this range of centralities, the measured yields are in good agreement with the predictions, while for mid-peripheral and peripheral collisions corresponding to \avgNpart\ $<200$, there is a slight excess of $W^\pm$ bosons in data in comparison with the theory predictions. The effect grows as \avgNpart\ decreases. It is largest  in the most peripheral bin and amounts to 1.7~(0.8) standard deviations for $W^-$~($W^+$) boson production. After combining the two bins with the lowest  \avgNpart\ values, the excess in measured normalised production yields over the theory predictions is 1.7~(0.9) standard deviations for $W^-$~($W^+$) bosons. It was checked whether the events from the lowest \avgNpart\ bin could be contaminated by a contribution from photonuclear background. No significant enhancement of events with asymmetric signals in the ZDC on either side of ATLAS was seen.
 
\begin{figure}[!tbh]
\centering
\includegraphics[width=0.49\textwidth]{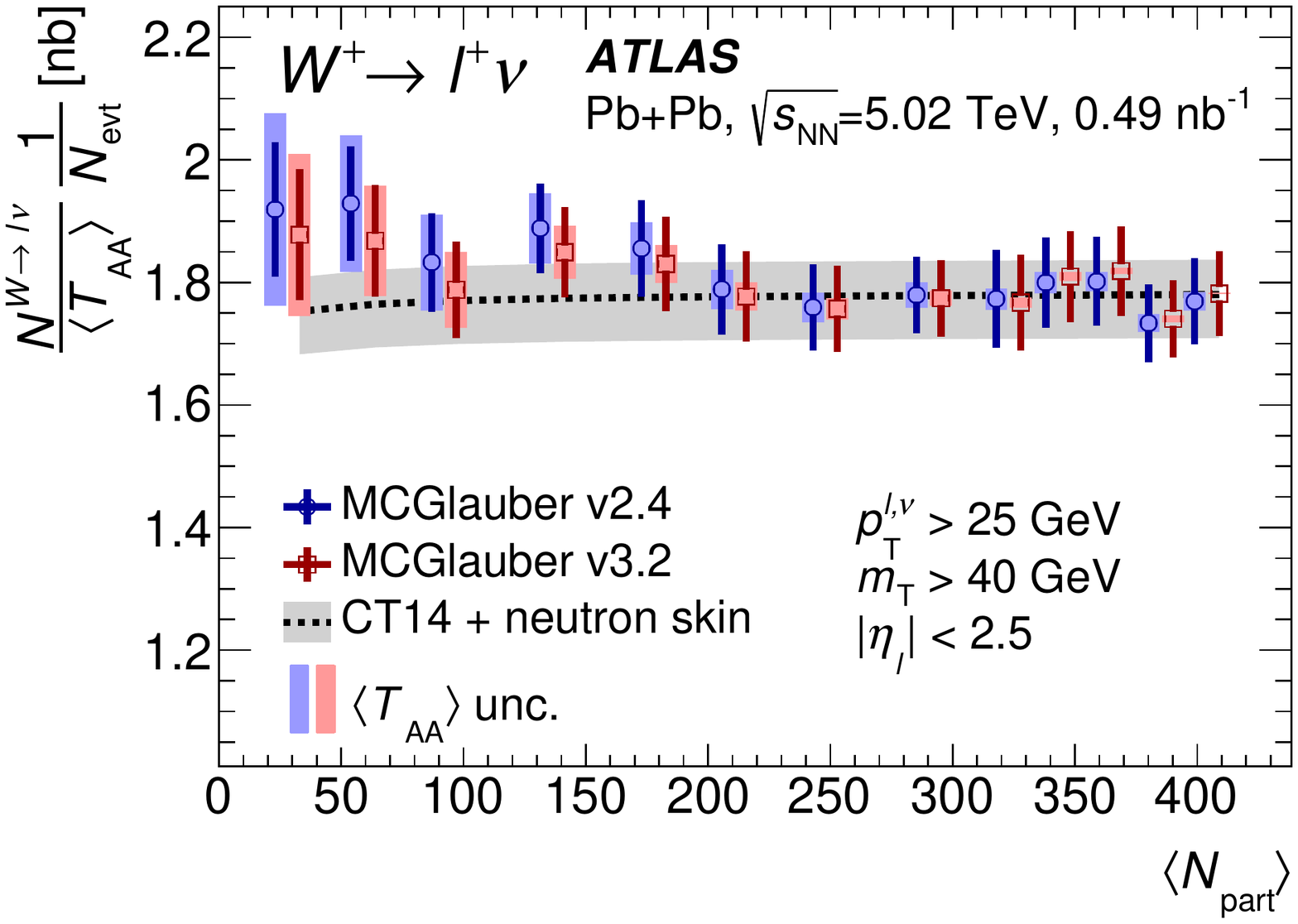}
\hfill
\includegraphics[width=0.49\textwidth]{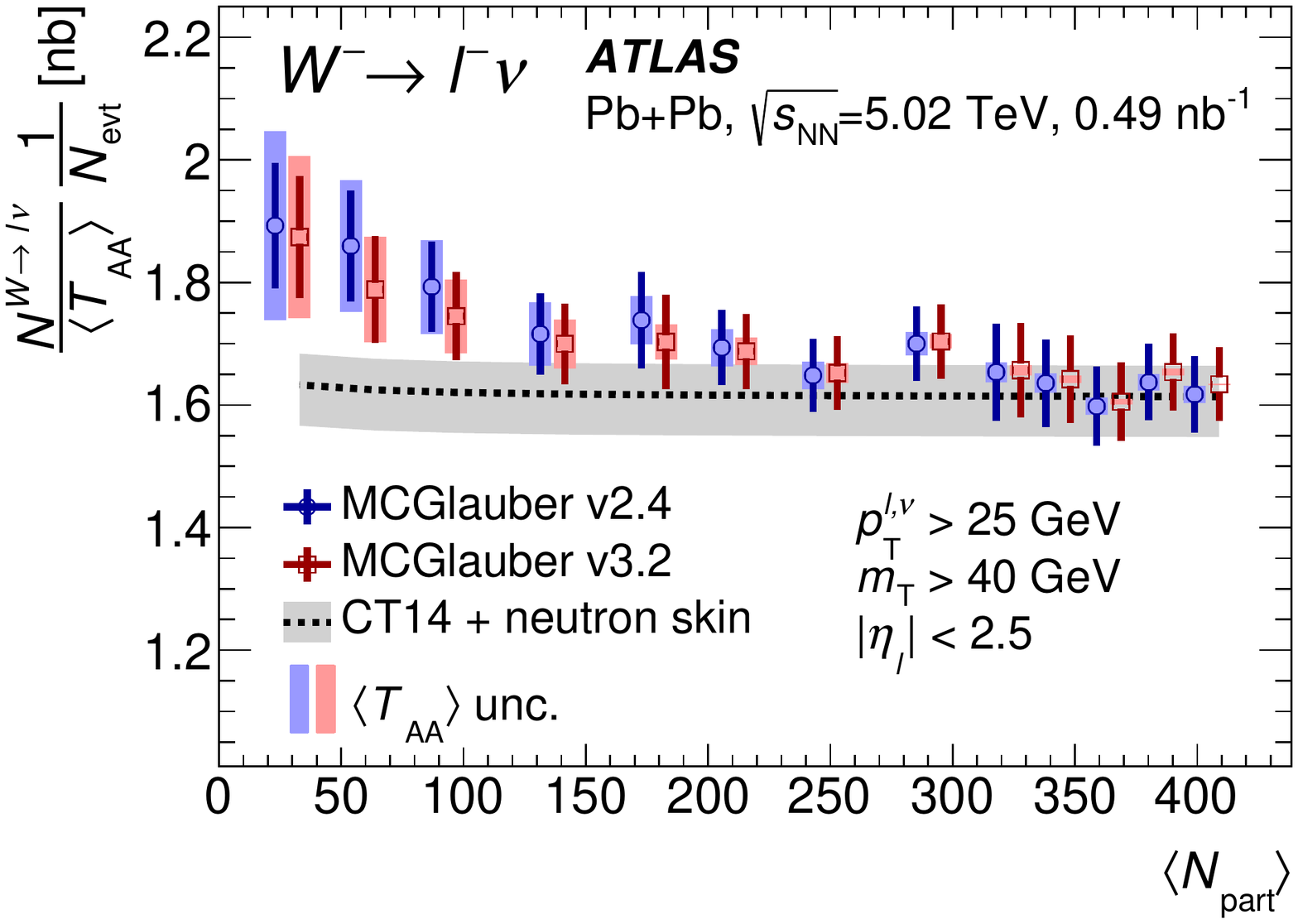}
\caption{Comparison of normalised production yields for $W^+$~(left) and $W^-$~(right) bosons as a function of \avgNpart\ for geometric parameters obtained with the MCGlauber code v2.4 and v3.2. The dashed lines show predictions calculated using the \textsc{CT14 NLO} PDF set which incorporate the neutron-skin effect evaluated using the separate radial distributions for protons and neutrons provided by the MCGlauber code v3.2. Uncertainties related to the determination of radial distributions for nucleons are not included. Error bars show statistical and systematic uncertainties added in quadrature, whereas systematic uncertainties due to \avTaa\ are shown as shaded boxes around the data points. The points for the MCGlauber code v3.2 are shifted horizontally for better visibility.}
\label{fig:glauber-comparison}
\end{figure}

The measurement of normalised production yields for $W^+$ and $W^-$ bosons is repeated using the alternative FCal~\sumet\ ranges to define centrality classes, \Nevt, \avgNpart\ and \avTaa\ values, extracted from the MCGlauber code v3.2. The results obtained using the two different MCGlauber  codes are compared in Figure~\ref{fig:glauber-comparison}. For both the $W^+$ and $W^-$ bosons, the normalised production yields extracted with geometric parameters from the MCGlauber code v3.2 are slightly closer to the constant yields expected from a scaling with the nuclear thickness. This improvement is more pronounced in peripheral events, but the MCGlauber code v3.2 results still do not fully follow a constant scaling. In addition, differences between the yields obtained using the MCGlauber code v2.4 and v3.2 are smaller than the experimental uncertainties. Theoretical predictions shown in Figure~\ref{fig:glauber-comparison} are calculated using the \textsc{CT14 NLO} PDF set and incorporate the neutron-skin effect~\cite{Paukkunen:2015bwa} evaluated using the separate radial distributions for protons and neutrons provided by the MCGlauber code v3.2. The difference between the radial distributions results in an evolution of the effective proton-to-neutron ratio with centrality. The impact of the neutron skin on normalised $W^\pm$ boson production yields is largest in the most peripheral collisions, where the predictions differ by $-1.4\%$ ($+1\%$) for $W^+$~($W^-$) bosons relative to predictions calculated using a constant proton-to-neutron ratio.
 
\begin{figure}[!tbh]
\centering
\includegraphics[width=0.49\textwidth]{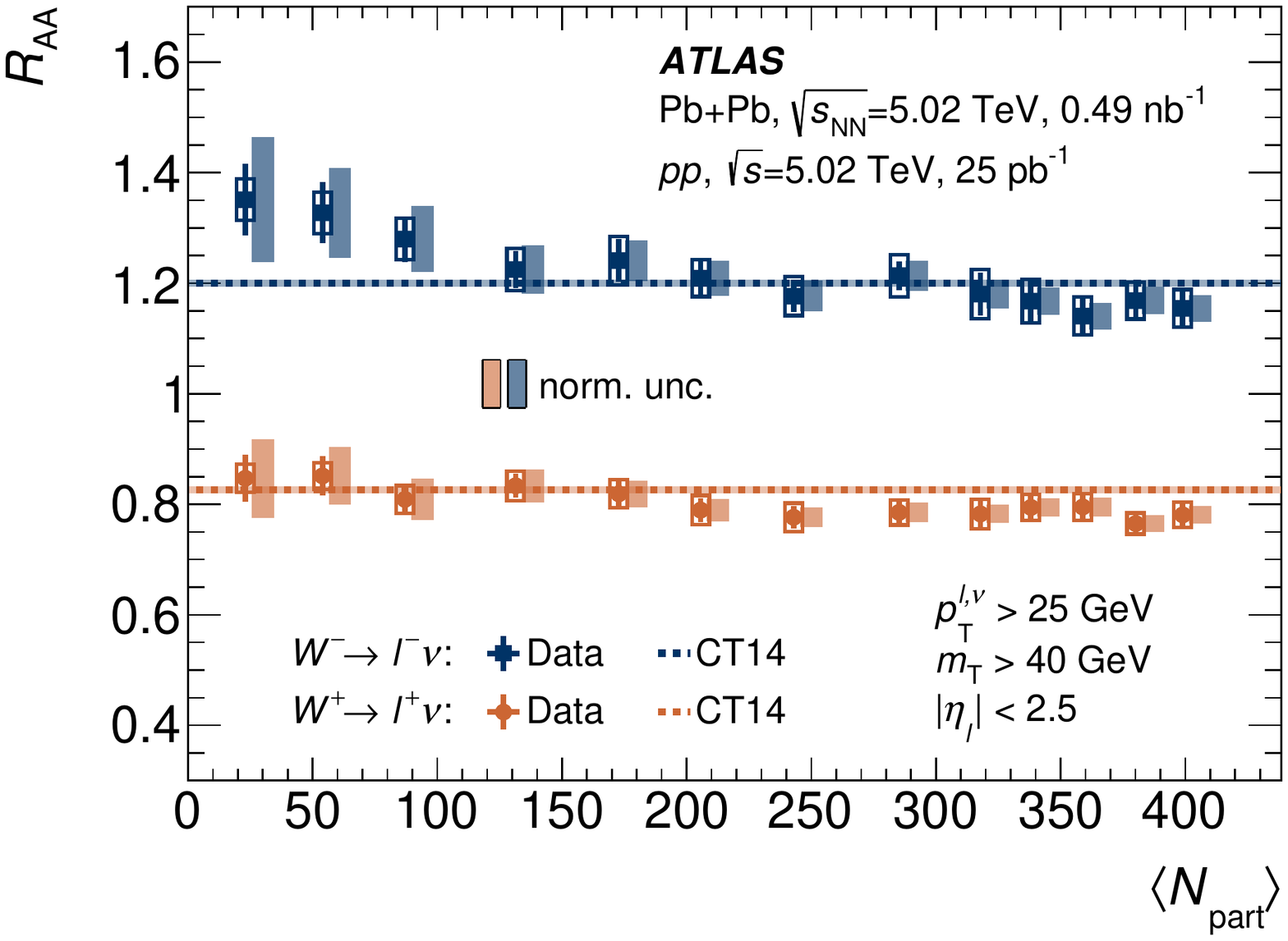}
\caption{Nuclear modification factor \Raa\ obtained from the fiducial $W^+$ and $W^-$ boson production yields as a function of \avgNpart.
Error bars show statistical uncertainties, whereas systematic uncertainties are shown as the boxes around the data points. The systematic uncertainties due to \avTaa\ and luminosity of the \pp\ dataset are not included in those boxes, and are shown as separate shaded boxes plotted to the right of the data points for better visibility.}
\label{fig:result-Raa}
\end{figure}

Figure~\ref{fig:result-Raa} shows the nuclear modification factor defined via Eq.~\eqref{eq:raa} as a function of \avgNpart\ for the production of $W^+$ and $W^-$ bosons for the combined electron and muon channels.
The \pp\ measurements used to obtain the \Raa\ factor come from Ref.~\cite{HION-2018-02-witherratum}. All uncertainties are assumed to be uncorrelated between the measurements in the \PbPb\ and \pp\ systems, and, therefore, are added in quadrature. As a function of \avgNpart, the nuclear modification factors for both the $W^+$ and $W^-$ bosons follow the same trend as the normalised production yields. The observed deviations of \Raa\ from unity can be mostly attributed to the isospin effect present in the \PbPb\ system, which results in an enhancement of $W^-$ bosons and a suppression of $W^+$ bosons relative to the \pp\ system. These modifications of $W^\pm$ boson production in the \PbPb\ system arise from the larger fraction of valence $d$-quarks in lead nuclei than in protons, since the dominant production mode of $W^\pm$ bosons is through $u\bar{d} \to W^+$ and $d\bar{u} \to W^-$ processes. The measured \Raa\ factors are compared with theoretical predictions calculated with the \textsc{CT14 NLO} PDF set. These predictions do not fully describe the \Raa\ factors despite reproducing the normalised production yields of $W^+$ and $W^-$ bosons measured as a function of $|\eta_\ell|$. For peripheral collisions, the measured \Raa\ factors agree with predictions within 1.2 (0.4) standard deviations for $W^-$~($W^+$) bosons, while for central collisions the agreement is within 1.1 (1.8) standard deviations. The apparent contradiction in the theoretical description of \Raa\ factors and of the  normalised production yields shown in Figure~\ref{fig:result-centrality} is due to the $W^\pm$ boson production cross-sections measured in the \pp\ system~\cite{HION-2018-02-witherratum} being larger than the CT14 NLO theory predictions.
 
\begin{figure}[!tbh]
\centering
\includegraphics[width=0.49\textwidth]{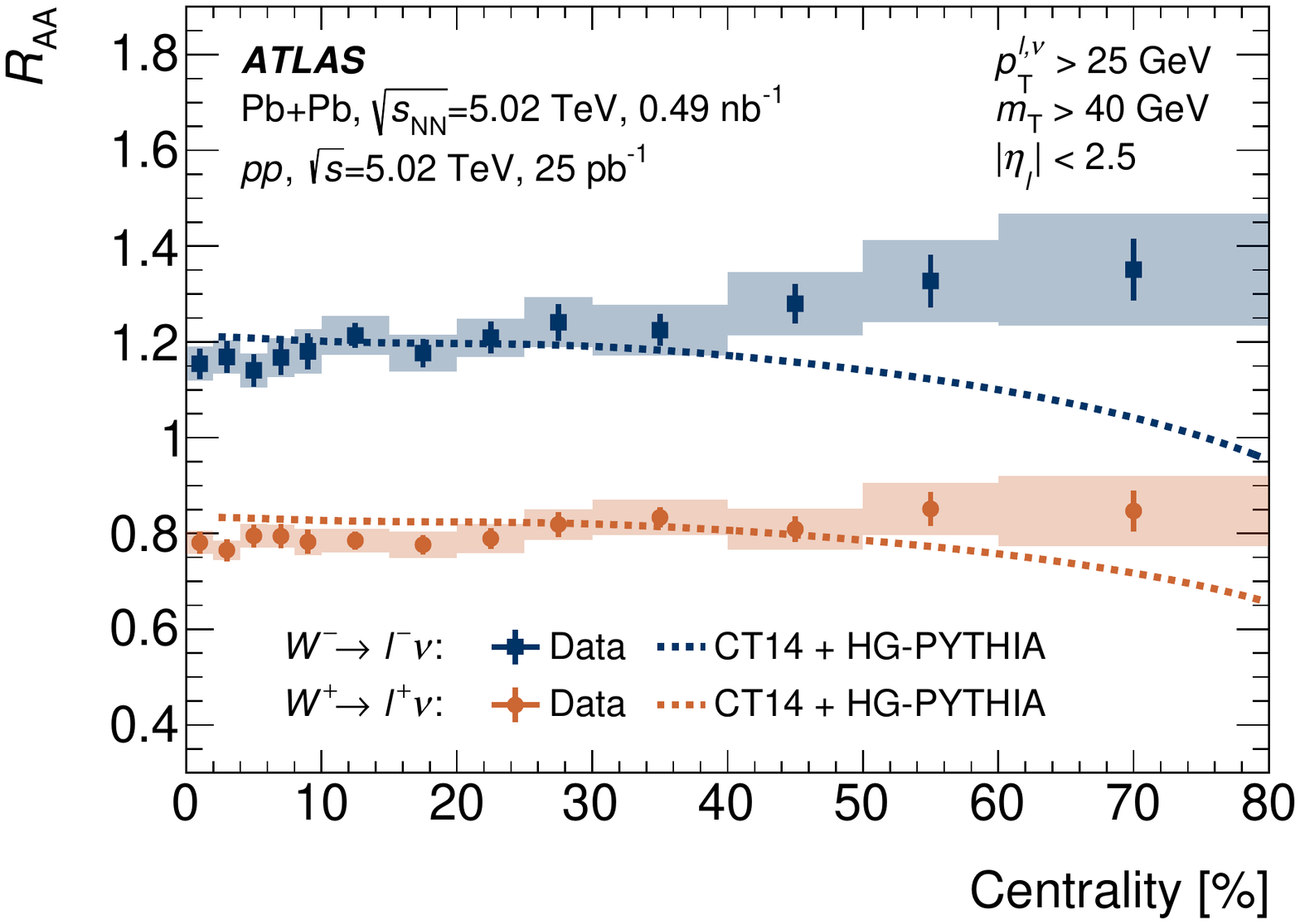}
\caption{Nuclear modification factor \Raa\ obtained from the fiducial $W^+$ and $W^-$ boson production yields as a function of centrality percentile. Error bars show statistical uncertainties, whereas systematic uncertainties are shown as the boxes around the data points. The systematic uncertainties due to \avTaa\ and luminosity of the \pp\ dataset are included in those boxes. The dashed lines show predictions which incorporate the centrality bias calculated using HG-PYTHIA for the ALICE charged-hadron production measurement~\cite{Acharya:2018njl}. These predictions are corrected for the isospin effect evaluated using MCFM with the CT14 NLO PDF set.}
\label{fig:result-Raa-centrality}
\end{figure}

It was recently argued in Ref.~\cite{Morsch:2017brb} that the value of \Raa\ in peripheral A+A collisions is expected to be modified due to a biased classification of the event geometry, as specified by the centrality assignment, for events containing a hard process. In that analysis, the value of \Raa\ without any nuclear effects was determined using the HG-PYTHIA model, which describes A+A collisions using the \textsc{Hijing} event generator~\cite{Wang:1991hta} and superimposes a \PYTHIA\ 6.4 event~\cite{Sjostrand:2006za-correct} for each hard sub-interaction specified by the generator. The HG-PYTHIA prediction was demonstrated to describe the ALICE measurement of charged-hadron \Raa\ in peripheral Pb+Pb collisions~\cite{Acharya:2018njl}. In Figure~\ref{fig:result-Raa-centrality}, the \Raa\ factors for $W^+$ and $W^-$ bosons measured as a function of centrality percentile are compared with theoretical predictions calculated with the \textsc{CT14 NLO} PDF set multiplied by the HG-PYTHIA prediction of the \Raa\ factor taken from Ref.~\cite{Acharya:2018njl}. The impact of re-evaluating the bias for centrality classes defined in the ATLAS centrality determination procedure is found to be negligible. Due to the different interplay of the hard process and soft production for events with jet production and EW boson production, it is not clear that the centrality bias in this analysis is expected to be the same as in the ALICE measurement. In fact, the observed trend of \Raa\ factors with centrality goes in the opposite direction to the effect of centrality bias predicted by HG-PYTHIA for charged-hadron production.

\clearpage
% End of text imported from the .//Sec_5.tex input file
% The next lines are included from the .//Sec_6.tex input file
\section{Summary and conclusions}
\label{sec:summary}
Inclusive production of $W^\pm$ bosons decaying into $\ell^\pm\nu$ measured in the electron and muon channels in \PbPb\ collisions at $\sqrtNN = 5.02$~\TeV\ is reported. The measurements are based on the data collected in 2015 using the ATLAS detector at the LHC corresponding to an integrated luminosity of 0.49~nb$^{-1}$.
 
The fiducial production yields scaled by the average nuclear thickness function, \avTaa, and the total number of minimum-bias \PbPb\ collisions, $N_\mathrm{evt}$, are measured in the phase-space region defined by the charged-lepton transverse momentum, $\pt^{\ell}>25$~\GeV, and pseudorapidity, $|\eta_\ell|<2.5$, the transverse momentum of the (anti)neutrino, $\pt^\nu>25$~\GeV, and the transverse mass of the charged-lepton--(anti)neutrino system, $\mt>40$~\GeV.
 
The dominant background contribution comes from multi-jet production. It is evaluated using a data-driven method and amounts to up to about 20\% and 12\% in the electron and muon decay channels, respectively. Other, smaller, background contributions come from EW boson decays ($Z\to\mu^+\mu^-(\tau^+\tau^-)$ and $W^\pm\to\tau^\pm\nu$) and $t\bar{t}$ production. They are estimated by normalising MC simulations to the integrated luminosity of the data sample.
 
After background subtraction and efficiency corrections, the normalised production yields
are presented as a function of the absolute pseudorapidity of the charged lepton, $|\eta_\ell|$, and the average number of nucleons participating in the collision, \avgNpart.
 
The normalised production yields for $W^\pm$ bosons are compatible in the two  lepton decay channels which are combined in this analysis. The combined normalised production yields are consistent with theoretical predictions based on the \textsc{CT14 NLO} PDF set, while predictions obtained with the EPPS16 and nCTEQ15 nPDF sets underestimate the measured yields by 10--20\%. However, measurements in the \pp\ system have shown that \textsc{CT14 NLO} predictions underestimate the data by 4--5\%.
The measured yields for $W^\pm$ bosons are also used to obtain the lepton charge asymmetry, which is well described by the mentioned theoretical predictions. The lepton charge asymmetry changes sign and becomes negative for $|\eta_\ell| > 2$, which is an indication of the isospin effect yielding a larger fraction of $W^-\to\ell^-\nu$ events in \PbPb\ compared to \pp\ collisions. The data lack sensitivity to nuclear modifications of the nucleon PDFs in the kinematic region probed by $W^\pm$ boson production. Normalised production yields for $W^\pm$ bosons are in agreement with the expected scaling with \avTaa\  for mid-central and central events. In the range \avgNpart $<200$, a systematic excess of the normalised production yields of $W^\pm$ bosons is observed in the data in comparison with the theory predictions with the isospin effect included. The effect is largest in the two most peripheral bins for $W^-$ bosons, and there, the combined excess amounts to 1.7 standard deviations.
A comparison of normalised production yields for geometric parameters obtained with two versions of the MCGlauber code (v2.4 and v3.2) shows that the MCGlauber code v3.2 results are somewhat closer to the constant yields expected from the scaling with the nuclear thickness. However, the difference between the two results is smaller than the measurement uncertainties after the uncertainty in \avTaa\ is excluded.
 
Nuclear modification factors, \Raa, for $W^\pm$ boson production are calculated using cross-sections measured in \pp\ collisions at the same centre-of-mass energy.
As a function of \avgNpart, the measured \Raa\ factors follow the same trend as the normalised production yields. The observed deviations of \Raa\ from unity can be mostly attributed to the isospin effect. For peripheral collisions, the data agree with predictions based on the \textsc{CT14 NLO} PDF set within 1.2 (0.4) standard deviations for $W^-$~($W^+$) bosons, while for central collisions the agreement is within 1.1 (1.8) standard deviations. The trend observed in the data goes in the opposite direction to predictions including the centrality bias evaluated with the HG-PYTHIA model for charged-hadron production. Due to differences in the soft-particle production between events with jet production and EW boson production, it is not clear that the centrality bias should be the same for these two classes of processes. The presented measurement can be used to constrain the centrality bias in the future.

% End of text imported from the .//Sec_6.tex input file
 
\clearpage
\section*{Acknowledgements}
% The next lines are included from the .//acknowledgements/Acknowledgements.tex input file
 
We thank CERN for the very successful operation of the LHC, as well as the
support staff from our institutions without whom ATLAS could not be
operated efficiently.
 
We acknowledge the support of ANPCyT, Argentina; YerPhI, Armenia; ARC, Australia; BMWFW and FWF, Austria; ANAS, Azerbaijan; SSTC, Belarus; CNPq and FAPESP, Brazil; NSERC, NRC and CFI, Canada; CERN; CONICYT, Chile; CAS, MOST and NSFC, China; COLCIENCIAS, Colombia; MSMT CR, MPO CR and VSC CR, Czech Republic; DNRF and DNSRC, Denmark; IN2P3-CNRS, CEA-DRF/IRFU, France; SRNSFG, Georgia; BMBF, HGF, and MPG, Germany; GSRT, Greece; RGC, Hong Kong SAR, China; ISF and Benoziyo Center, Israel; INFN, Italy; MEXT and JSPS, Japan; CNRST, Morocco; NWO, Netherlands; RCN, Norway; MNiSW and NCN, Poland; FCT, Portugal; MNE/IFA, Romania; MES of Russia and NRC KI, Russian Federation; JINR; MESTD, Serbia; MSSR, Slovakia; ARRS and MIZ\v{S}, Slovenia; DST/NRF, South Africa; MINECO, Spain; SRC and Wallenberg Foundation, Sweden; SERI, SNSF and Cantons of Bern and Geneva, Switzerland; MOST, Taiwan; TAEK, Turkey; STFC, United Kingdom; DOE and NSF, United States of America. In addition, individual groups and members have received support from BCKDF, CANARIE, CRC and Compute Canada, Canada; COST, ERC, ERDF, Horizon 2020, and Marie Sk{\l}odowska-Curie Actions, European Union; Investissements d' Avenir Labex and Idex, ANR, France; DFG and AvH Foundation, Germany; Herakleitos, Thales and Aristeia programmes co-financed by EU-ESF and the Greek NSRF, Greece; BSF-NSF and GIF, Israel; CERCA Programme Generalitat de Catalunya, Spain; The Royal Society and Leverhulme Trust, United Kingdom.
 
The crucial computing support from all WLCG partners is acknowledged gratefully, in particular from CERN, the ATLAS Tier-1 facilities at TRIUMF (Canada), NDGF (Denmark, Norway, Sweden), CC-IN2P3 (France), KIT/GridKA (Germany), INFN-CNAF (Italy), NL-T1 (Netherlands), PIC (Spain), ASGC (Taiwan), RAL (UK) and BNL (USA), the Tier-2 facilities worldwide and large non-WLCG resource providers. Major contributors of computing resources are listed in Ref.~\cite{ATL-GEN-PUB-2016-002}.
 
% End of text imported from the .//acknowledgements/Acknowledgements.tex input file
\printbibliography

\clearpage % ATLAS Collaboration author list
% Reference date of HION-2018-25 is 2019-04-24
% Author list last updated on date 14-NOV-19
% Data extracted on 14-Nov-2019 for paper reference HION-2018-25
% at 8:52am
 
\begin{flushleft}
{\Large The ATLAS Collaboration}

\bigskip

G.~Aad$^\textrm{\scriptsize 101}$,    
B.~Abbott$^\textrm{\scriptsize 128}$,    
D.C.~Abbott$^\textrm{\scriptsize 102}$,    
A.~Abed~Abud$^\textrm{\scriptsize 70a,70b}$,    
K.~Abeling$^\textrm{\scriptsize 53}$,    
D.K.~Abhayasinghe$^\textrm{\scriptsize 93}$,    
S.H.~Abidi$^\textrm{\scriptsize 167}$,    
O.S.~AbouZeid$^\textrm{\scriptsize 40}$,    
N.L.~Abraham$^\textrm{\scriptsize 156}$,    
H.~Abramowicz$^\textrm{\scriptsize 161}$
% ...
\bigskip
\\

% ...
$^{40}$Niels Bohr Institute, University of Copenhagen, Copenhagen; Denmark.\\
% ...
$^{53}$II. Physikalisches Institut, Georg-August-Universit\"{a}t G\"ottingen, G\"ottingen; Germany.\\
% ...
$^{70}$$^{(a)}$INFN Sezione di Pavia;$^{(b)}$Dipartimento di Fisica, Universit\`a di Pavia, Pavia; Italy.\\
% ...
$^{93}$Department of Physics, Royal Holloway University of London, Egham; United Kingdom.\\
% ...
$^{101}$CPPM, Aix-Marseille Universit\'e, CNRS/IN2P3, Marseille; France.\\
$^{102}$Department of Physics, University of Massachusetts, Amherst MA; United States of America.\\
% ...
$^{128}$Homer L. Dodge Department of Physics and Astronomy, University of Oklahoma, Norman OK; United States of America.\\
% ...
$^{156}$Department of Physics and Astronomy, University of Sussex, Brighton; United Kingdom.\\
% ...
$^{161}$Raymond and Beverly Sackler School of Physics and Astronomy, Tel Aviv University, Tel Aviv; Israel.\\
% ...
$^{167}$Department of Physics, University of Toronto, Toronto ON; Canada.\\
% ...
$^{a}$ Also at Borough of Manhattan Community College, City University of New York, New York NY; United States of America.\\
$^{b}$ Also at CERN, Geneva; Switzerland.\\
% ...

\end{flushleft}

% Created with Glance <Atlas.Glance@cern.ch>

\end{document}